\documentclass[sigconf]{acmart}
\AtBeginDocument{%
  }

\setcopyright{acmlicensed}
\copyrightyear{2026}
\acmYear{2026}
\acmDOI{XXXXXXX.XXXXXXX}

\acmConference[XXX '27]{XXXX}{XXX 19—23, 2027}{XXX, XXX}  
\acmISBN{978-1-4503-XXXX-X/2018/06}

\usepackage{enumitem}
\usepackage{graphicx}
\usepackage{textcomp}
\usepackage{xcolor}
\usepackage[normalem]{ulem}
\usepackage{bbding}
\usepackage{pifont}
\usepackage{amsmath,amsfonts,amsthm}
\usepackage[linesnumbered, ruled, vlined]{algorithm2e}
\usepackage{algpseudocode}
\usepackage{booktabs}
\usepackage{tabularx}
\usepackage{multirow}
\usepackage{setspace}
\usepackage{enumitem}
\usepackage{microtype}
\usepackage{subcaption}
\fontfamily{unihei}

\begin{document}

\title{LizTail: Unifying Shape Adaptation and Spatial Parallelism with Segment-Reconfigurable Systolic Arrays}

\author{Gang Liu}
\authornote{Both authors contributed equally to this research.}
\email{gliu29@uestc.edu.cn}
\orcid{0000-0003-1330-0945}
\author{Chen Yang}
\authornotemark[1]
\email{202622280624@std.uestc.edu.cn}
\affiliation{%
  \institution{Shenzhen Institute for Advanced Study, University of Electronic Science and Technology of China}
  \city{Shenzhen}
  \country{China}
}

\author{Junyan Hu}
\email{junyanhu@xupt.edu.cn}
\affiliation{%
  \institution{Xi'an University of Posts and Telecommunications}
  \city{Xi'an}
  \country{China}}

\author{Jianguo Chen}
\email{chenjg33@mail.sysu.edu.cn}
\affiliation{%
  \institution{School of Software Engineering, Sun Yat-sen University}
  \city{Zhuhai}
  \country{China}
}

\author{Lixin Duan}
\email{lxduan@uestc.edu.cn}
\affiliation{%
 \institution{Shenzhen Institute for Advanced Study, University of Electronic Science and Technology of China}
 \city{Shenzhen}
 \country{China}}





\renewcommand{\shortauthors}{XXX et al.}

\begin{abstract}
Systolic arrays are widely used to accelerate neural-network inference, yet fixed two-dimensional arrays struggle with the substantial diversity in both matrix geometry and available parallelism across CNNs and Transformer-based large language models. This mismatch is particularly severe for skinny matrices, small-batch autoregressive decoding, and operators containing multiple independent partitions, leading to low processing-element (PE) utilization and repeated spatial folding. Existing reconfigurable arrays typically improve either shape adaptation or distributed execution, whereas supporting both through fine-grained routing incurs considerable interconnect, buffering, and control overheads.
We present LizTail\footnote{Named after lizard tail autotomy, where predefined fracture planes allow segments to detach and remain independently active, analogous to LizTail's predefined reconfiguration boundaries.}, a multi-mode systolic array that enables constrained segment-level reconfiguration by partitioning a $128\times128$ array into 32 regular $128\times4$ segments and adding connectivity only at predefined endpoints.
The resulting substrate supports trimming, long-path extension through $180^\circ$ turn-back connections, and distributed execution in which each subarray can itself be trimmed or extended. Differentiated PE datapaths and segment-aware CEB/SIB buffers provide the required compute and data-delivery flexibility without unrestricted PE-level routing. A hardware-constrained mapper selects the execution organization, dataflow, task partition, and tiling.
Across representative workloads, LizTail achieves geometric-mean speedups of $5.02\times$ on CNN inference and $25.67\times$ on LLM decoding over a TPU-like array, while reducing EDP by up to $38.23\times$.
LizTail is open-sourced at \url{https://github.com/kmjbwxa/LizTail}.
\end{abstract}





\maketitle

\section{Introduction}

Matrix multiplication dominates modern neural-network inference, yet its shapes have become increasingly diverse~\cite{park2025dots, wang2025desa, lee2024resa, han2024redas, ye2025flashinfer, lee2025aif}. CNNs expose substantially different GEMM aspect ratios across layers~\cite{ramirez2024parallel, lee2024resa, pao2025eda}, while
Transformers and large language models (LLMs) introduce even greater heterogeneity across prefill and autoregressive decoding~\cite{vaswani2017attention, radford2019language,llama3,yang2025qwen3}. In particular, small-batch decoding contains many skinny GEMMs and GEMV-like operations in which one output dimension is much smaller than the others. Such workloads map poorly to large fixed two-dimensional systolic arrays, leaving a substantial fraction of processing elements (PEs) structurally underutilized.

Prior work~\cite{genc2021gemmini, ghodrati2020planaria,han2024redas, samajdar2022self} explores complementary points in this design space. DyNNamic~\cite{hanson2022dynnamic} adapts a monolithic array under a fixed dataflow, whereas ReDas~\cite{han2024redas} chains four physical regions through PE-local roundabout paths to construct fine-grained elongated logical arrays under OS, WS, or IS. Planaria~\cite{ghodrati2020planaria} exposes coarse-grained array fission, while SARA~\cite{samajdar2022self} supports variable array dimensions and numbers of subarrays through cell-granular scale-up and scale-out. 
These designs establish the benefits of shape adaptation, dataflow flexibility, and spatial partitioning, but differ substantially in workload scope, supported organizations, and data-delivery cost.
Meanwhile, our analysis of GEMM/GEMV layer shapes and their execution behavior in CNN and LLM inference tasks reveals a key finding: no single array organization consistently delivers optimal performance across various workloads (as shown in Figures \ref{fig:layer-breakdown} and \ref{fig:fixed-extend-parallel}).

This motivates a key question: \textbf{\emph{Can an accelerator capture both shape adaptation and spatial parallelism without paying the cost of unrestricted reconfiguration?}}

We present \textbf{LizTail}, a multi-mode systolic-array architecture based on constrained segment-level reconfiguration. \textbf{LizTail} partitions a $128\times128$ array into 32 regular $128\times4$ segments while preserving conventional systolic connectivity within each segment. Additional connectivity is introduced only at predefined endpoints. This substrate supports three complementary organizations: trimming disables unused segments; extension connects segments through $180^\circ$ turn-back paths to construct logical dimensions larger than the physical array; and distributed execution partitions a workload across multiple subarrays. Importantly, each parallel subarray may itself be trimmed or extended, allowing spatial parallelism and local shape adaptation to be composed.


\textbf{LizTail} co-designs compute reconfiguration with data delivery. Differentiated PE datapaths provide additional routing only where needed, while CEBs and SIBs provide segment-aware operand delivery without requiring a fully flexible global operand-delivery fabric. A hardware-constrained mapper selects the execution mode, logical shape, dataflow, task partition, and tiling for each workload.

Across representative CNNs and LLMs, LizTail achieves geometric-mean speedups of $5.02\times$ and $25.67\times$, respectively, over a TPU-like fixed array. On the evaluated LLMs, it further outperforms ReDas by $2.72$--$3.14\times$ and reduces EDP by up to $38.23\times$ over the fixed-array baseline.

The main contributions of this work are as follows:
\begin{itemize}[noitemsep,nolistsep,topsep=0pt]
\item We show that shape extension and spatial parallelism occupy complementary performance regimes across diverse GEMM/GEMV inference workloads.
\item We introduce a segment-reconfigurable systolic substrate, differentiated PE datapaths, and segment-aware CEB/SIB buffering that compose these capabilities with bounded hardware cost.
\item We develop a hardware-constrained mapper and demonstrate substantial performance, energy, and area-efficiency benefits across CNN and LLM inference.
\end{itemize}

\section{Workload Characterization and Design Motivation}
\label{WorkloadCharacterizationandDesignMotivation}
Modern neural-network inference exposes substantial diversity not only in matrix dimensions but also in the forms of parallelism available across operators. This section characterizes these workload properties and examines how they interact with different systolic-array organizations. 
We first summarize why fixed systolic-array geometry is sensitive to workload dimensions and dataflow \cite{jia2014caffe, chetlur2014cudnn, kwon2023efficient, vaswani2017attention, hong2024flashdecoding++, hanson2022dynnamic, aminabadi2022deepspeed}, as demonstrated for compact CNN layers and general DNN GEMMs~\cite{simonyan2014very, vgg16_geeksforgeeks, hanson2022dynnamic,han2024redas, krizhevsky2012imagenet, koonce2021resnet, he2016deep, tan2019efficientnet, kansal2024resnet, ren2015faster}. We then quantify shape heterogeneity across representative CNN and LLM workloads, including the regular GEMMs of prefill and the skinny GEMMs and GEMV-like operations prevalent in small-batch decoding.
We next compare fixed, shape-extended, and distributed-parallel organizations under an equal PE budget to determine whether a single organization can efficiently cover these workloads. Finally, we discuss why simply maximizing reconfiguration flexibility is undesirable and derive the architectural requirements that motivate LizTail.

\subsection{Workload Shape Diversity}
 \begin{figure}[htbp]
    \centering
    \includegraphics[width=0.95\linewidth]{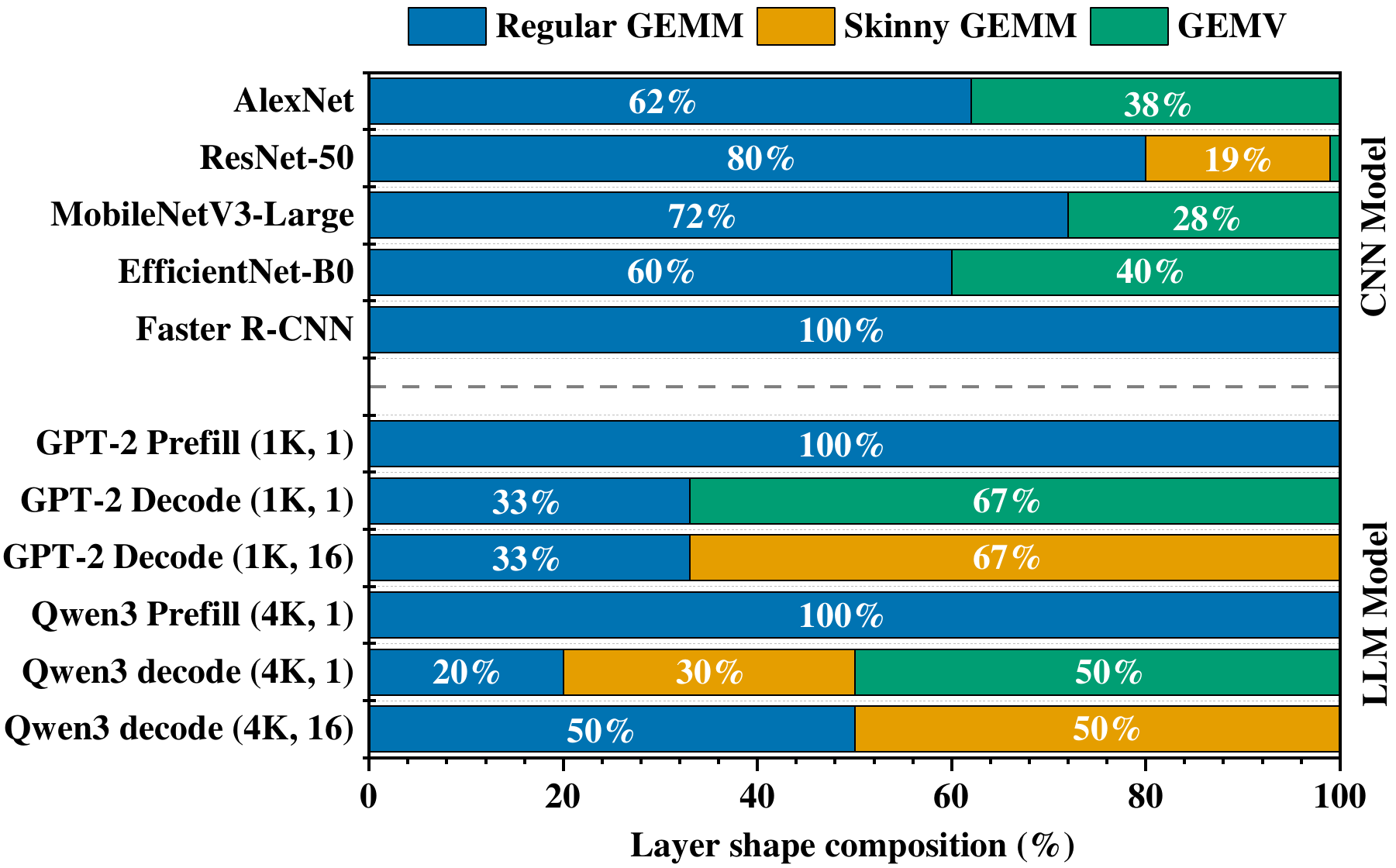}
    \caption{Distribution of regular GEMM, skinny GEMM, and GEMV operations across representative CNN and LLM workloads. Small-batch LLM decoding contains a substantially higher fraction of skinny GEMM/GEMV workloads than prefill and conventional CNN inference.}
    \label{fig:layer-breakdown}
\end{figure}


For a GEMM \(C_{M\times N}=A_{M\times K}B_{K\times N}\), mapping efficiency on an \(R\times C\) systolic array depends strongly on the relationship between the workload and array dimensions. A fixed geometry leaves PEs idle when a mapped dimension is smaller than the array and introduces repeated spatial folds when it exceeds the array dimension~\cite{hanson2022dynnamic,han2024redas}. Compact CNN layers further alter data-reuse opportunities, causing their preferred array aspect ratios to vary across layer types~\cite{hanson2022dynnamic}.

Figure~\ref{fig:layer-breakdown} quantifies this diversity across representative CNN \cite{krizhevsky2012imagenet, koonce2021resnet, he2016deep, tan2019efficientnet, kansal2024resnet, ren2015faster} and LLM \cite{openai2022chatgpt, yang2025qwen3} workloads. For characterization, we classify $M=1$ as GEMV, $1<M\leq\min(N,K)/8$ as skinny GEMM, and all remaining operations as regular GEMM. CNNs contain a mixture of regular and skinny GEMMs, whereas small-batch LLM decoding contains substantially more skinny GEMM/GEMV operations. Prefill and larger-batch decoding shift back toward regular two-dimensional GEMMs.

\textbf{Observation 1.} \emph{Inference does not expose a single dominant GEMM/\\GEMV shape; the preferred array geometry varies across models, layers, and inference phases.
A fixed array optimized for one dominant matrix geometry therefore cannot maintain consistently high utilization across this workload spectrum.}

\subsection{Why No Single Array Organization Suffices}
We next compare three organizations under the same 16K-PE budget: a fixed $128\times128$ array, a $32\times512$ extended array, and four $32\times128$ parallel subarrays. Each configuration considers both orientations and OS/WS/IS dataflows.

\begin{figure}[htbp]
    \centering
    \includegraphics[width=0.9\linewidth]{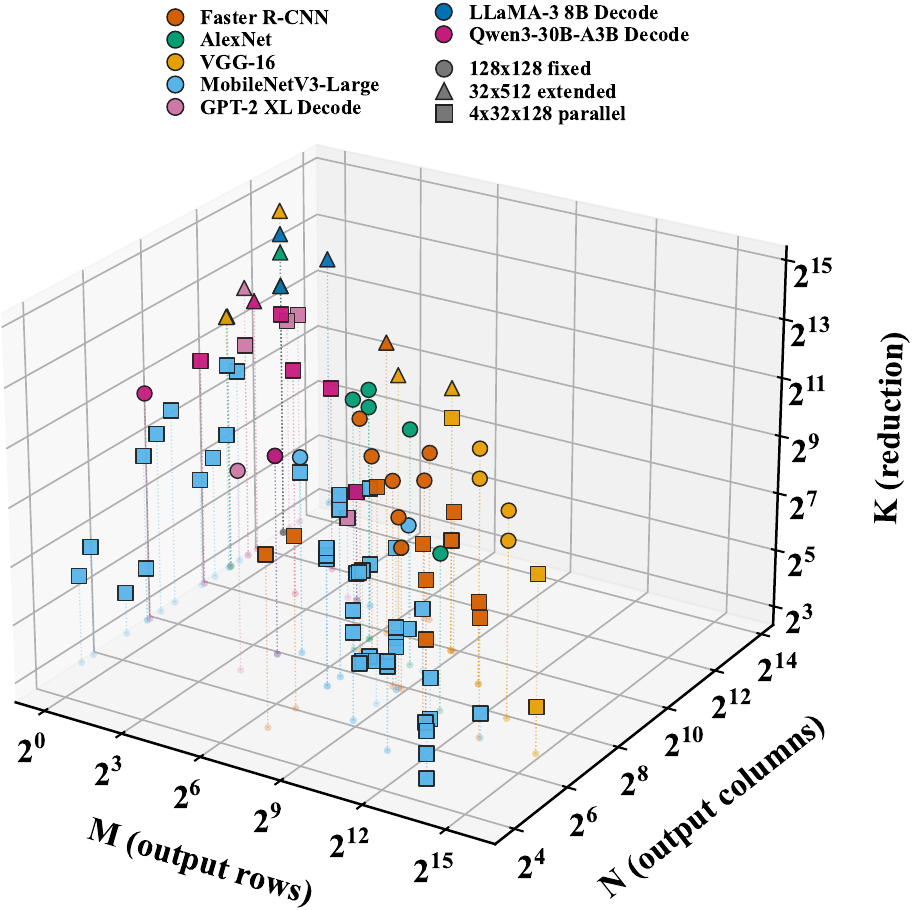}
    \caption{Preferred array organizations across representative CNN and LLM GEMM/GEMV workloads under an equal PE budget. Fixed, shape-extended, and distributed-parallel organizations dominate in different regions of the workload space, demonstrating that no single organization consistently performs best.}
    \label{fig:fixed-extend-parallel}
\end{figure}

Figure~\ref{fig:fixed-extend-parallel} shows that the preferred organization varies substantially across workloads. Elongated GEMM/GEMV operations favor extension because a longer logical dimension
reduces folding. Partitionable workloads instead benefit from multiple subarrays. For example, partitioning the $(M,N,K)=(1,4800,1600)$ GPT-2 XL \cite{openai_gpt2xl_hf} QKV projection along $N$ across four $32\times128$ subarrays reduces cycles by $17.9\%$ relative to a single $32\times512$ extended array. Large,
well-balanced GEMMs can still favor the native array.

\textbf{Observation 2.} \emph{Shape extension and spatial parallelism alleviate different bottlenecks and are complementary rather than interchangeable. No single array organization consistently dominates across diverse GEMM/GEMV workloads.}

\subsection{Why Not Fully Flexible Reconfiguration?}
A broadly reconfigurable topology must provide both flexible compute paths and sufficient operand delivery. Global turns and source-to-PE bypasses can introduce long wires, multiplexing, and highly banked buffers. ReDas~\cite{han2024redas} demonstrates a more scalable alternative for shape adaptation: it constructs roundabout paths using only bidirectional links between adjacent PEs. This avoids global bypass wiring, but requires four-directional ports, crossbars, and forwarding registers throughout the PE array, together with four banked multi-mode edge buffers. Planaria~\cite{ghodrati2020planaria} and SARA~\cite{samajdar2022self} occupy different points in this design space through array fission and cell-granular scale-up/scale-out, respectively. These designs establish the value of reconfiguration while illustrating that its cost depends on where routing and operand-delivery flexibility are instantiated.

Our characterization shows that useful mappings concentrate in four recurring organizations: native/trimmed arrays, elongated paths, independent subarrays, and parallel subarrays with local shape adaptation. Supporting arbitrary PE topologies is therefore unnecessary.

\noindent\textbf{Observation 3.} \textit{Workload-relevant mappings occupy a small, structured subset of the arbitrary PE-topology space.}

\begin{figure}[htbp]
    \centering
    \includegraphics[width=0.95\linewidth]{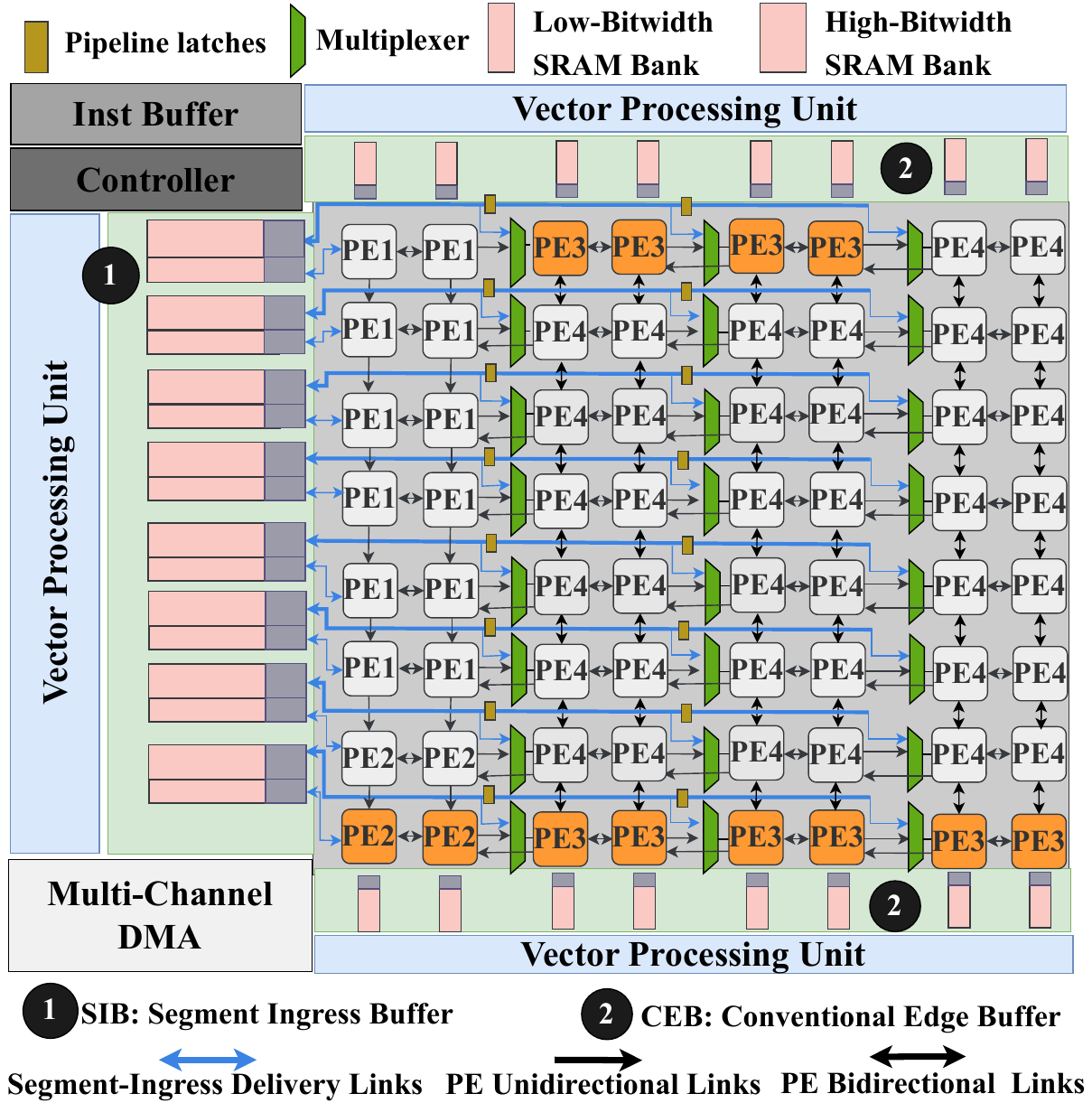}
    \caption{Overview of LizTail. The evaluated design contains a \(128\times128\) PE array partitioned into 32 \(128\times4\) segments; a four-segment \(8\times8\) example is shown for clarity. LizTail combines constrained inter-segment connectivity, differentiated PE datapaths, and segment-aware CEB/SIB buffering. }
    \label{fig:arch-overview}
\end{figure}

\section{LizTail Architecture}
\label{sec:architecture}
Section \ref{WorkloadCharacterizationandDesignMotivation} shows that efficient inference requires more than a single form of array reshaping: elongated workloads benefit from extending the logical execution path, while partitionable workloads benefit from spatial parallelism, and some workloads require both. At the same time, supporting arbitrary PE-level reconfiguration would introduce excessive routing and data-delivery complexity. Guided by these observations, LizTail adopts a segment-reconfigurable substrate that restricts reconfiguration to regular array segments and predefined inter-segment endpoints.

The central principle is simple: preserve conventional systolic connectivity within a segment, and introduce additional flexibility only where it is required to compose segments into useful logical organizations. This section first presents the overall architecture and its three execution modes, and then describes the inter-segment connectivity, differentiated PE datapaths, and segment-aware SRAM organization that realize these modes.

\subsection{Architecture Overview}
\begin{figure*}[!htbp]
    \centering
    \includegraphics[width=0.95\textwidth]{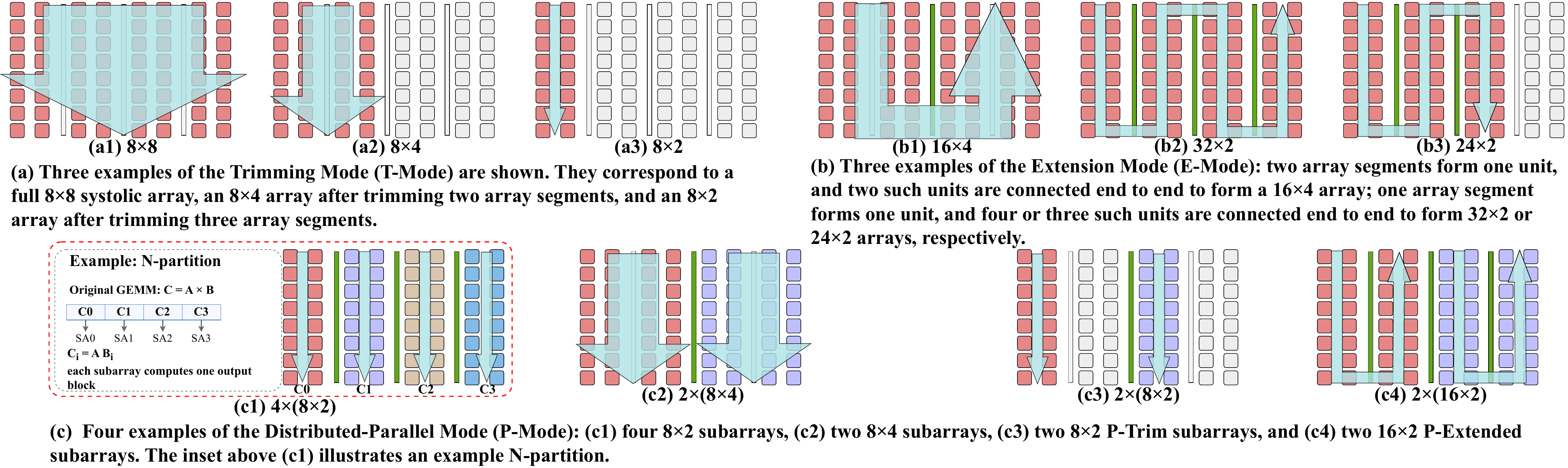}
    \caption{LizTail execution modes on the illustrative \(8\times8\) array: (a) T-Mode reduces the active array footprint, (b) E-Mode concatenates segments into elongated logical arrays, and (c) P-Mode executes multiple partitions concurrently and can further combine with local trimming or extension.}
    \label{fig:three-modes}
\end{figure*}

Figure~\ref{fig:arch-overview} presents the LizTail architecture. The evaluated design partitions a \(128\times128\) PE array into 32 regular \(128\times4\) segments; a scaled-down \(8\times8\) array with four \(8\times2\) segments is shown to expose the datapaths and inter-segment connections. LizTail comprises four main components: a segment-reconfigurable PE array, segment-aware on-chip buffers, vector-processing units, and a control and data-movement subsystem.

The PE array preserves conventional neighbor-to-neighbor connectivity within each segment and introduces additional routing only at predefined segment boundaries and endpoints. By configuring the active segments and their logical connectivity, the same substrate supports native or trimmed arrays, elongated logical paths, and multiple concurrently executing subarrays. 
Section~\ref{Multi_Mode_Execution} describes these execution modes, Section \ref{sec:segment_connect} details the endpoint connections and differentiated PE datapaths, while Section \ref{sec:sram_buffer} describes the segment-aware data-delivery organization.

Operand delivery is co-designed with compute reconfiguration. Conventional Edge Buffers (CEBs) retain conventional boundary-based data delivery, whereas Segment Ingress Buffers (SIBs) inject operands only at predefined segment ingress points. This organization supplies extended and distributed configurations without requiring a fully flexible global operand-delivery fabric that can route data from on-chip buffers to arbitrary PEs. 
The vector-processing units execute element-wise operations and reduce partial sums generated by~\(K\)-partitioned workloads.

The LizTail Mapper selects the execution mode, logical array shape, dataflow, task partition, tiling, and buffer placement offline. At runtime, the controller programs the selected segment configuration, while the multi-channel DMA transfers operands between off-chip memory and the corresponding CEB/SIB banks. This separation keeps runtime control simple while exposing workload-specific array organizations.

\begin{figure}[!htbp]
    \centering
    \includegraphics[width=0.5\textwidth]{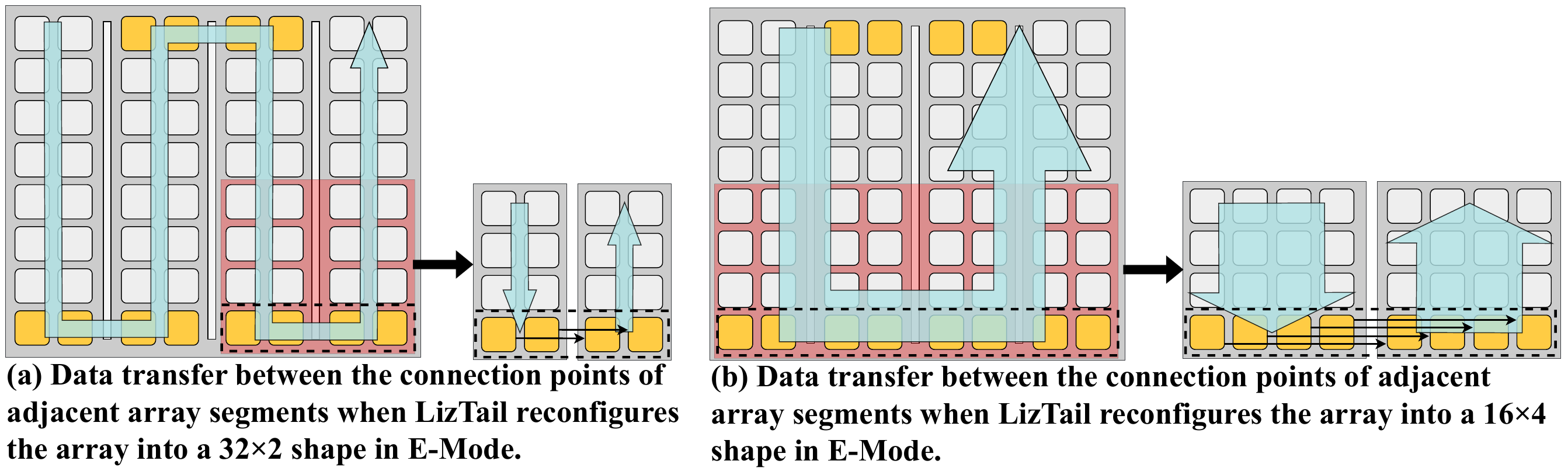}
    \caption{Inter-segment routing for elongated logical arrays. (a,b) LizTail uses predefined endpoint-based \(180^\circ\) turn-back connections to construct \(32\times2\) and \(16\times4\) logical arrays.}
    \label{fig:segment-connection}
\end{figure}
\subsection{Multi-Mode Execution}
\label{Multi_Mode_Execution}

Figure~\ref{fig:three-modes} illustrates how LizTail organizes the same physical array into three execution modes. These modes differ in both the number of active segments and their logical connectivity, allowing the architecture to cover the recurring workload regimes identified in Section~\ref{WorkloadCharacterizationandDesignMotivation}.

\subsubsection{Trimming Mode (T-Mode)}
T-Mode (as shown in Figure \ref{fig:three-modes}(a)) targets workloads whose useful spatial footprint is smaller than the physical array. It activates (S) contiguous ($128\times4$) segments to form a ($128\times4S$) logical array, or its transpose when permitted by the selected dataflow, and disables the remaining segments. Because T-Mode preserves conventional adjacency within the active region, it requires no turn-back connections. Trimming reduces the propagation boundary and suppresses switching in unused PEs and interconnects, but cannot construct a logical dimension beyond the physical array boundary.

\subsubsection{Extension Mode (E-Mode)}
E-Mode (as shown in Figure \ref{fig:three-modes}(b)) targets elongated GEMM/GEMV workloads whose preferred logical dimension exceeds the physical side length. E-Mode reconnects complete segments through legal ($180^{\circ}$) turn-back endpoints, forming an elongated path while retaining regular systolic propagation within each segment. The connected segments collectively execute a single GEMM/GEMV rather than independent workload partitions. This longer effective dimension reduces spatial folding and the associated repeated mapping and fill/drain overheads. Section~3.3 describes the endpoint connections that realize these paths.



\subsubsection{Distributed-Parallel Mode (P-Mode).}
P-Mode targets GEMM/\\GEMV workloads that can be partitioned across multiple concurrently executing subarrays.
Unlike E-Mode, which uses multiple segments to form a single logical systolic array, P-Mode first divides the physical array into multiple independently executing subarrays and assigns a different workload partition to each subarray. We distinguish two types of parallelism: \emph{output-space partitioning} along $M$ or $N$, and \emph{reduction-space partitioning} along $K$.
Consider
$\mathbf{C}_{M\times N}
=
\mathbf{A}_{M\times K}
\mathbf{B}_{K\times N}.$

An $M$-partition divides $\mathbf{A}$ and $\mathbf{C}$ by rows: each subarray computes $\mathbf{C}_i=\mathbf{A}_i\mathbf{B}$ for a disjoint subset of output rows. Similarly, an $N$-partition divides $\mathbf{B}$ and $\mathbf{C}$ by columns:
$\mathbf{B}
=
[\mathbf{B}_1|\cdots|\mathbf{B}_P],
\qquad
\mathbf{C}
=
[\mathbf{C}_1|\cdots|\mathbf{C}_P],$
where each subarray independently computes $\mathbf{C}_i=\mathbf{A}\mathbf{B}_i$. Because $M$- and $N$-partitions produce disjoint output regions, they require no cross-subarray partial-sum reduction.

$N$-partitioning is particularly important for small-batch LLM decoding, where $M$ is often one and therefore provides little or
no partitioning opportunity. For example, the GPT-2 XL QKV projection with $(M,N,K)=(1,4800,1600)$ can be divided across four subarrays by assigning 1,200 output columns to each subarray. All four subarrays consume the same input vector $\mathbf{A}_{1\times1600}$ but different weight blocks $\mathbf{B}_i\in\mathbb{R}^{1600\times1200}$, and produce four independent output blocks that are concatenated to form the final result. This is the output-dimension parallelism used in the motivation example of Section~\ref{WorkloadCharacterizationandDesignMotivation}.

In contrast, a $K$-partition splits the reduction dimension:
$\mathbf{A}
=
[\mathbf{A}_1|\cdots|\mathbf{A}_P],
\qquad
\mathbf{B}
=
[\mathbf{B}_1^T|\cdots|\mathbf{B}_P^T]^T .$
Each subarray then produces a partial result $\mathbf{C}_i^{\mathrm{partial}}=\mathbf{A}_i\mathbf{B}_i$, and the final output requires $\mathbf{C}=\sum_i\mathbf{C}_i^{\mathrm{partial}}$. LizTail performs this cross-subarray reduction using the vector
processing unit, and the Mapper includes the associated reduction and synchronization costs when evaluating $K$-partitioned
candidates.

Figure~\ref{fig:three-modes}(c) illustrates the \emph{physical subarray organizations} rather than a specific $M/N/K$ partition. After choosing the workload partition, each subarray can independently adopt one of three local organizations: \emph{P-Native} preserves its assigned geometry, \emph{P-Trim} disables unnecessary local segments, and \emph{P-Extended} connects its local segments into an elongated path. Thus, P-Mode separates two decisions---\emph{what portion of the GEMM each subarray computes} and \emph{how that portion is mapped locally}---allowing spatial parallelism to be composed with shape adaptation.

\begin{figure}[htbp]
    \centering
    \includegraphics[width=0.95\linewidth]{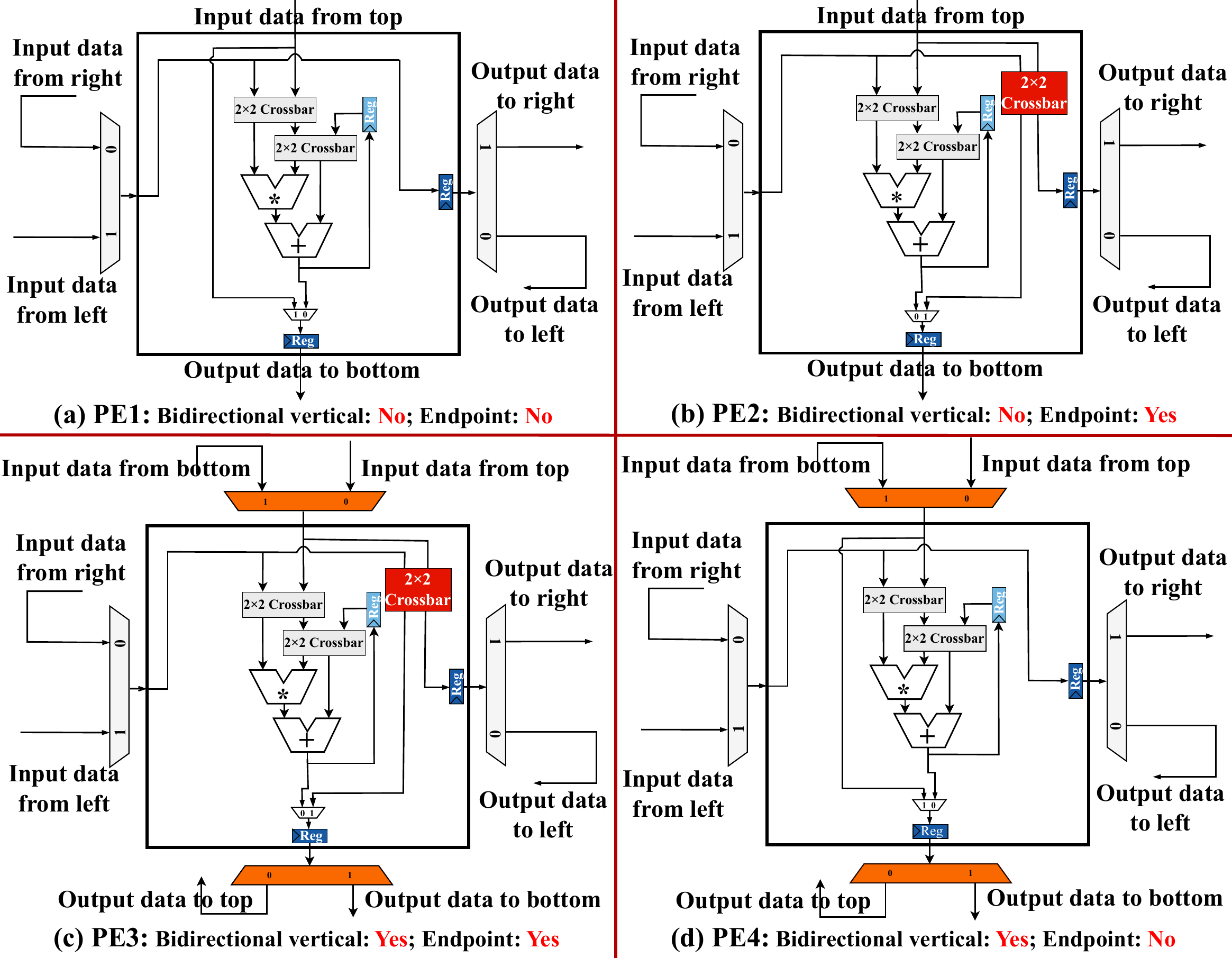}
    \caption{Differentiated PE datapaths. PE1–PE4 are derived from two routing requirements: vertical bidirectionality and participation in a segment endpoint connection.}
    \label{fig:pe-types}
\end{figure}

\subsection{Segment-Level Reconfiguration}
\label{sec:segment_connect}
LizTail realizes E-Mode without a general inter-PE routing network. As shown in Figure~\ref{fig:segment-connection}, cross-segment transitions occur only at predefined endpoints. Data traverses one segment, crosses horizontally at its endpoint, and then propagates in the opposite vertical direction through the adjacent segment, forming a serpentine $180^\circ$ turn-back path. Segment-internal connectivity therefore remains identical to a conventional systolic array.

ReDas~\cite{han2024redas} also constructs elongated logical arrays by chaining four variable-size regions through roundabout paths. Its design enables fine-grained reshaping by equipping every PE with bidirectional ports, crossbars, and pass-through capability; some PEs therefore relay operands rather than perform useful MACs. 
Unlike ReDas~\cite{han2024redas}, LizTail adopts a different reconfiguration trade-off.
It preserves conventional connectivity within each \(128\times4\) segment and instantiates turn-back connections only at predefined segment endpoints. More importantly, the same segments can either form one elongated path or operate as multiple concurrent subarrays, each of which can be locally trimmed or extended. LizTail’s contribution is therefore not turn-back routing alone, but the constrained composition of long-path extension and spatial parallelism.

This organization introduces only two additional routing capabilities—bidirectional vertical propagation and endpoint connectivity—whose four combinations define PE1--PE4, as shown in Figure~\ref{fig:pe-types}.
All variants share the same MAC datapath; only MUX/DEMUX and forwarding logic differ. Consequently, the most flexible datapath is instantiated only where required by legal segment paths rather than replicated across the entire array.

\subsection{Segment-Aware Data Delivery}
\label{sec:sram_buffer}
Changing the logical array organization also changes where operands must enter the array. LizTail therefore combines Conventional Edge Buffers (CEBs) with Segment Ingress Buffers (SIBs). CEBs retain
conventional boundary delivery, whereas SIBs inject operands only at predefined segment ingress points, avoiding a fully flexible operand-delivery fabric to arbitrary PEs.

For the four-segment example, supplying one INT8 operand per segment requires 32 bits of aggregate ingress bandwidth. LizTail splits this into an 8-bit bank serving the leftmost segment and a
24-bit bank serving the remaining segments. Small trimmed configurations therefore activate only the narrow bank, while extended or distributed organizations enable the wider path.
Pipeline latches break long SIB paths to maintain the target clock frequency.

\begin{figure*}[htbp]
    \centering
    \includegraphics[width=0.95\textwidth]{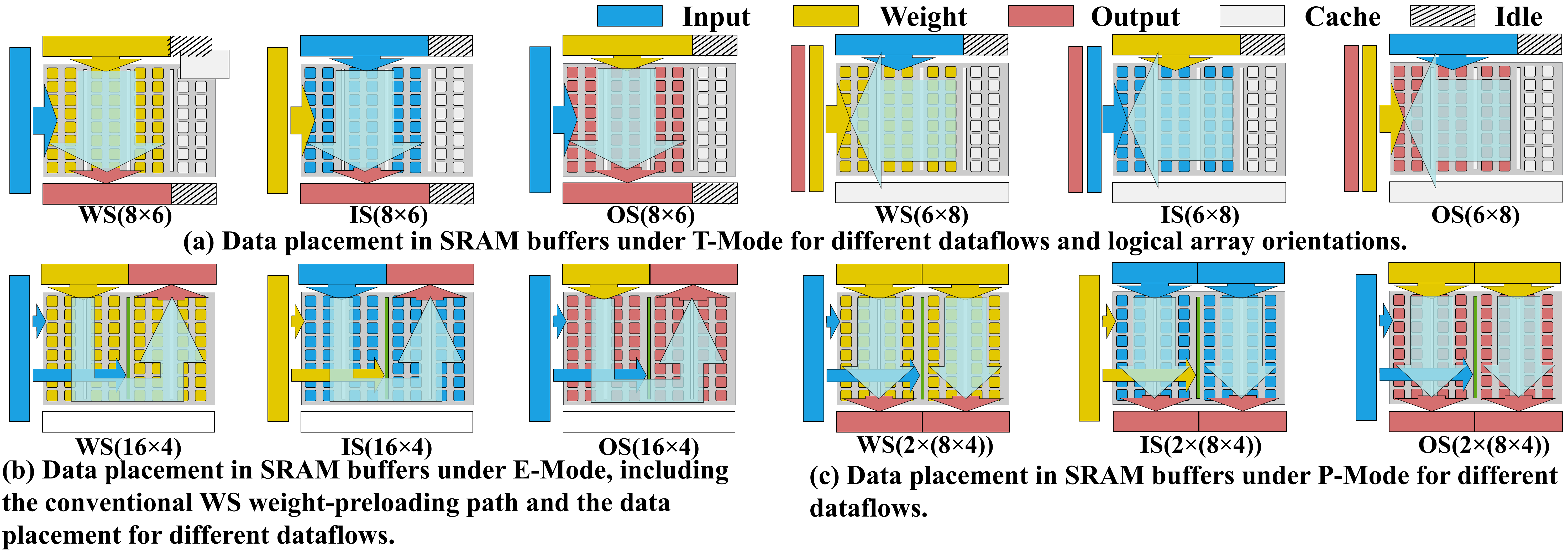}
    \caption{CEB/SIB data placement under different execution modes and dataflows. Segment-aware operand injection decouples physical data delivery from the logical array organization.}
    \label{fig:sram-buffer-data-placement}
\end{figure*}

Figure~\ref{fig:sram-buffer-data-placement} shows representative placement under T/E/P-Mode. Physical data delivery is deliberately decoupled from logical topology: segments can be preloaded through
their physical ingress points before forming an elongated path, and multiple P-Mode subarrays can be supplied concurrently. $K$-partitioned execution forwards partial sums to the vector unit
for reduction.

\section{LizTail Mapper}
\label{sec:mapper}
Compared with the ReDas Mapper~\cite{han2024redas} design, LizTail targets a broader execution hierarchy that additionally includes trimming, task partitioning, multiple concurrent subarrays, and local extension within each partition. Its search reduction therefore derives from segment-constrained candidate generation, independent T/E/P-Mode branches, critical-boundary tiling, and repeated-shape caching rather than interval sampling alone.

Section~\ref{sec:architecture} provides LizTail with a bounded but expressive configuration space: segments can be selectively activated, connected into elongated paths, or partitioned into multiple concurrently executing subarrays. The remaining challenge is to identify an efficient configuration for each GEMM/GEMV workload. This task is non-trivial because the best configuration depends jointly on the execution mode, logical array geometry, dataflow, task partitioning, tiling, loop ordering, and operand placement. These choices are further constrained by the legal segment connections, CEB/SIB capacity and bandwidth, off-chip memory bandwidth, and reduction overheads introduced by parallel execution.

A naive mapper could enumerate the Cartesian product of all these choices and evaluate every configuration. Such an approach, however, quickly becomes impractical even after physically unrealizable array shapes are removed. LizTail therefore follows a different principle: \emph{hardware constraints are applied before software mapping choices are expanded}. The mapper first constructs only those logical organizations that can be realized by the segmented substrate, and then explores dataflow, partitioning, tiling, and loop ordering within this legal configuration space.

\begin{figure}[!htbp]
  \centering
  \includegraphics[width=0.95\linewidth]{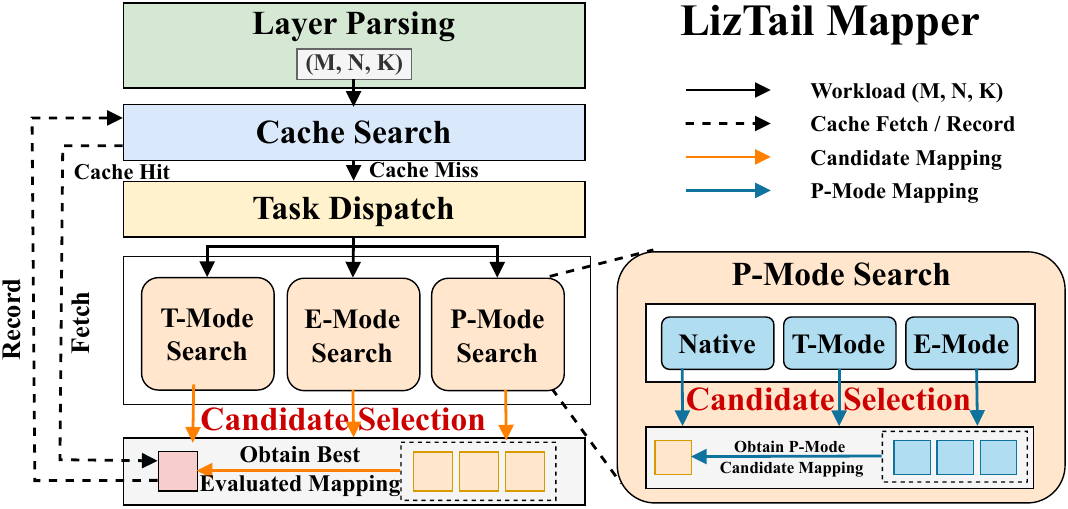}
  \caption{Overview of the LizTail Mapper. The mapper first lowers neural-network operators into GEMM/GEMV tasks and queries the repeated-workload cache. On a cache miss, hardware constraints derived from the segmented architecture restrict the legal T-Mode, E-Mode, and P-Mode organizations; P-Mode further explores P-Native, P-Trim, and P-Extended local configurations. The mapper then expands each legal organization over dataflows, buffer-feasible placements, critical tilings, and loop orders, evaluates all candidates using a unified latency/energy cost model, and selects the configuration that minimizes the target objective.}
  \label{fig:liztail-mapper}
\end{figure}

Figure~\ref{fig:liztail-mapper} summarizes the mapping flow. Neural-network operators are first lowered into GEMM/GEMV tasks. The mapper then queries a repeated-workload cache. On a cache miss, T/E/P-Mode candidates are generated under the architectural constraints defined in Section~\ref{sec:architecture}. P-Mode further considers P-Native, P-Trim, and P-Extended organizations. Each legal organization is expanded over the remaining mapping choices and evaluated using a unified latency/energy cost model. Finally, Candidate Selection chooses the configuration that minimizes the target optimization objective.
The entire mapping procedure is performed offline; only the selected configuration is communicated to the runtime controller.

\subsection{Mapping Problem and Mapper Overview}
\label{sec:mapper_overview}
Consider a GEMM workload 
$\mathbf{C}_{M\times N}
    =
    \mathbf{A}_{M\times K}
    \mathbf{B}_{K\times N}.$  
We represent a mapping candidate as
\begin{equation}
    x =
    \langle
    \mu,\,
    (H_{a},W_{a}),\,
    d,\,
    \pi,\,
    \tau,\,
    \ell
    \rangle ,
\end{equation}
where $\mu$ denotes the execution mode, $(H_{a},W_{a})$ is the logical array shape, $d$ is the dataflow, $\pi$ specifies the task partition and parallelism degree, $\tau=(M_t,N_t,K_t)$ denotes the tile shape, and $\ell$ specifies the loop order. Operand placement is generated only from the placements permitted by the selected execution mode, array orientation, and dataflow, and is subsequently checked against the CEB/SIB capacity and bandwidth constraints.

For workload $w=(M,N,K)$ and hardware configuration $\mathcal{H}$, let $\Omega(w,\mathcal{H})$ denote the set of physically realizable mapping candidates. The mapper solves
\begin{equation}
    x^{*}
    =
    \arg\min_{x\in\Omega(w,\mathcal{H})}
    J(x),
    \label{eq:mapper_objective}
\end{equation}
where $J(x)$ can represent execution latency, energy consumption, or energy-delay product (EDP). Unless otherwise stated, we use $ J(x) = E(x)\times T(x)$ as the default objective.

The key distinction between LizTail Mapper and an unconstrained mapping search lies in how $\Omega$ is constructed. LizTail does not first enumerate arbitrary PE rectangles and subsequently discard invalid configurations. Instead, the segment granularity and endpoint connectivity described in Section~\ref{sec:segment_connect} directly define the legal logical-array candidates. Software-level mapping choices are expanded only after this structural filtering. This hardware-first organization substantially reduces the number of candidates that require performance and energy evaluation.

\subsection{Segment-Constrained Candidate Generation}
\label{sec:candidate_generation}
LizTail constructs the candidate space hierarchically. The mapper first generates legal logical organizations for each execution mode and then expands each organization over task partitioning, dataflow, buffer placement, tiling, and loop ordering.

\paragraph{Trimming candidates.}
In T-Mode, the mapper activates a contiguous set of $S$ physical segments and disables the remaining segments, where $1 \leq S \leq 32$. Because each segment in the evaluated design has a physical size of $128\times4$, the canonical logical shape is $(H_{a}, W{a}) = (128,4S),$ with the transposed orientation also considered when permitted by the dataflow and operand-delivery constraints.

T-Mode does not modify the internal topology of the active segments.
Consequently, candidate generation only determines the number of active segments and their orientation before exploring the remaining
mapping choices. This makes T-Mode the most compact search branch.

\paragraph{Extension candidates.}
E-Mode constructs elongated logical arrays by connecting complete segments through the legal turn-back endpoints introduced in
Section~\ref{sec:segment_connect}. Importantly, the mapper does not enumerate arbitrary rectangles with the same number of PEs.
Suppose an E-Mode candidate uses $S$ segments and organizes them into $g$ parallel segment chains, where $g \mid S$.

Each chain contains $S/g$ segments connected through legal $180^{\circ}$ turn-back endpoints. Under the canonical orientation,
the resulting logical array satisfies
$  H_a = 4g,   
    W_a = 128\frac{S}{g}.$
The mapper also considers the transposed form $(W_a,H_a)$ when it is compatible with the selected dataflow and buffer organization.

For example, four segments can form $4\times512$, $8\times256$, or $16\times128$ canonical organizations (and their legal transposes), whereas shapes that cannot be constructed from complete segments and valid endpoint connections are discarded before subsequent mapping exploration. In this way, segment connectivity acts as a structural pruning rule rather than a post-processing legality check.

\paragraph{Distributed-parallel candidates.}
P-Mode introduces an additional hierarchy. The mapper first chooses the number of concurrently executing subarrays $   P \in \{2,4,8,16,32\},$ subject to workload and hardware feasibility. For an equal-size partition, each subarray receives $S_P = \frac{32}{P}$ physical segments.

Within this per-subarray segment budget, the mapper evaluates three local organizations: \emph{P-Native}, \emph{P-Trim}, and \emph{P-Extended}.
P-Native preserves the native rectangular organization assigned to each partition; P-Trim activates only a subset of its local segments; and P-Extended applies the same endpoint-constrained extension rule as E-Mode within each parallel partition. This hierarchy allows LizTail to compose spatial parallelism with local shape adaptation, rather than treating the two mechanisms as mutually exclusive.

For each parallel organization, the mapper further considers task partitioning along $M$, $N$, or $K$. Partitioning along $M$ or $N$
creates disjoint output regions and therefore does not require cross-subarray partial-sum reduction. In contrast, a $K$-partition
produces partial contributions to the same output tile. Such candidates are tagged with the corresponding reduction requirement so that the additional vector-processing cost can be included during evaluation.

\paragraph{Dataflow and buffer feasibility.}
For every legal array organization, the mapper evaluates 
$d \in \{\mathrm{OS},\mathrm{WS},\mathrm{IS}\}.$
The selected mode, logical-array orientation, and dataflow jointly determine the feasible CEB/SIB operand placements described in Section~\ref{sec:sram_buffer}. Placements that exceed the available buffer capacity, violate segment-ingress bandwidth, or cannot sustain the required concurrent subarrays are removed before cost evaluation.
Thus, buffer placement remains part of the mapping decision but is restricted to configurations realizable by LizTail's physical data delivery paths.

\paragraph{Critical tiling and loop ordering.}
After determining the mode, logical organization, and dataflow, the mapper must select the tile shape $\tau=(M_t,N_t,K_t)$. Exhaustively enumerating all integer-valued tile dimensions produces a large number of configurations with identical architectural behavior. LizTail therefore constructs a compact set of \emph{critical tiling candidates} around two classes of boundaries: (1) array-folding boundaries, where changing a tile dimension changes the number of spatial folds, and (2) CEB/SIB capacity boundaries, where a tile changes whether an operand can remain resident on chip. These boundaries capture the configurations at which execution cycles, data reuse, or memory-service behavior can change discontinuously.

For each retained tile configuration, the mapper evaluates the six permutations of the $M$-, $N$-, and $K$-tile loops. Different loop orders expose different reuse opportunities for inputs, weights, and partial outputs across consecutive tiles. By combining structural array pruning with critical tiling, LizTail avoids evaluating the large number of mappings that are either physically invalid or architecturally equivalent.

\subsection{Performance and Energy Cost Model}
\label{sec:cost_model}

All T/E/P-Mode candidates are evaluated using the same cost model. This is important because mode selection should be driven by the predicted end-to-end execution cost rather than by mode-specific heuristics such as PE utilization alone.
{For all architectures, a PE is counted as utilized only when performing a useful MAC; relay-only, disabled, and idle PEs remain part of the physical-PE denominator.}

\paragraph{Latency model.}
For candidate $x$, the predicted execution time is
\begin{equation}
    T(x)
    =
    \max
    \left\{
        T_{\mathrm{cmp}}(x),
        T_{\mathrm{mem}}(x)
    \right\}
    +
    T_{\mathrm{aux}}(x),
    \label{eq:latency_model}
\end{equation}
where $ T_{\mathrm{mem}}(x)
    =
    \max
    \left\{
        T_{\mathrm{buf}}(x),
        T_{\mathrm{DRAM}}(x)
    \right\}.$
$T_{\mathrm{cmp}}$ captures the systolic execution time determined by the logical array geometry, selected dataflow, tile count, spatial folding, and array fill/drain behavior.
$T_{\mathrm{buf}}$ represents the service time required by the CEB and SIB and accounts for active banks, bank width, operand placement, segment-ingress bandwidth, and the number of concurrently executing subarrays.
$T_{\mathrm{DRAM}}$ captures off-chip service time based on the traffic generated by the selected tiling and the available memory
bandwidth.

LizTail employs double buffering so that, during steady-state execution, most operand movement can overlap with systolic computation. Consequently, the critical steady-state path is approximated by the maximum of compute, on-chip data supply, and
off-chip memory service. $T_{\mathrm{aux}}$ captures costs that cannot be fully hidden by this overlap, including array
reconfiguration, turn-back-path overhead, parallel-task dispatch and synchronization, and the reduction latency required by
$K$-partitioned execution.

\paragraph{Energy model.}
The energy model follows the same candidate decomposition:
\begin{equation}
    E(x)
    =
    E_{\mathrm{PE}}(x)
    +
    E_{\mathrm{mem}}(x)
    +
    E_{\mathrm{route}}(x)
    +
    E_{\mathrm{par}}(x).
    \label{eq:energy_model}
\end{equation}

$E_{\mathrm{PE}}$ includes energy from useful MAC operations, activity in enabled but underutilized PEs, and PE leakage during the
candidate execution time.
$E_{\mathrm{mem}}$ includes CEB/SIB reads and writes, bank activation, buffer static energy, and off-chip DRAM accesses.
$E_{\mathrm{route}}$ accounts for additional communication along extension paths and pipelined segment-ingress delivery links.
Finally, $E_{\mathrm{par}}$ captures costs specific to distributed execution, including operand distribution, synchronization, and
partial-sum reduction when required.

The model derives buffer traffic from the selected operand placement, active array edges, bank widths, reuse pattern, and parallel sharing degree. Hence, a configuration with high nominal PE utilization is not automatically preferred if it increases memory traffic or parallelization overhead. This unified treatment allows the same objective in Eq.~\ref{eq:mapper_objective} to directly compare native, trimmed, extended, and distributed organizations.

\subsection{Mode-Parallel Search, Caching, and Selection}
\label{sec:parallel_search}
The candidate sets for T/E/P-Mode are mutually independent once the workload and hardware parameters are fixed.
LizTail therefore evaluates the three mode spaces using independent search branches. T-Mode Search explores trimming configurations,
E-Mode Search evaluates legal elongated paths, and P-Mode Search enumerates the supported parallelism degrees and evaluates
P-Native, P-Trim, and P-Extended local organizations.

Let the wall-clock search times of the three branches be $t_T$, $t_E$, and $t_P$. When sufficient host parallelism is available, mode-level parallel search reduces the serial search time from approximately
$ t_T+t_E+t_P$ to $\max\{t_T,t_E,t_P\}   +    t_{\mathrm{merge}},$ where $t_{\mathrm{merge}}$ denotes the small cost of merging and ranking evaluated candidates. Importantly, this optimization does
not remove candidates or change the optimization objective; it only exploits the independence among the three mode-specific searches to reduce mapper wall-clock time.

LizTail further exploits the repeated structure of neural networks. CNNs and, particularly, Transformer-based models frequently contain multiple layers or blocks with identical GEMM/GEMV dimensions. The mapper therefore maintains a workload cache indexed by $  k =
    \langle
      \mathrm{op},
      M,N,K,
      \mathcal{H},
      BW,
      J
    \rangle ,$ where $\mathrm{op}$ identifies the operator type, $\mathcal{H}$ captures the relevant hardware and candidate-generation parameters, $BW$ represents the configured memory bandwidth, and $J$ denotes the optimization objective. On a cache hit, the mapper reuses the previously evaluated candidate set instead of repeating the T/E/P searches.

After all mode-specific candidates have been evaluated, Candidate Selection merges the candidate sets and returns the configuration
with the minimum objective value according to Eq.~\ref{eq:mapper_objective}. The resulting configuration specifies the execution mode, logical array shape, dataflow, parallel partition, tile shape, loop order, and CEB/SIB placement required by
the runtime controller.

Overall, LizTail Mapper reduces mapping overhead through three orthogonal mechanisms. Hardware-constrained candidate generation
reduces the number of configurations that must be considered; mode-level parallelism reduces the wall-clock time required to
evaluate the remaining candidates; and repeated-workload caching avoids redundant searches across structurally identical layers.
Together, these mechanisms convert the expressive multi-mode hardware substrate described in Section~\ref{sec:architecture} into
automatic, workload-dependent execution decisions without resorting to unrestricted exhaustive search.

\section{Evaluation}
\label{sec:evaluation}
We evaluate LizTail to answer five questions. First, how much performance and energy-efficiency improvement does
LizTail provide over fixed and existing reconfigurable systolic arrays? Second, where do these gains come from, and are shape
extension and spatial parallelism both necessary? Third, what hardware cost does segment-level reconfiguration introduce?
Fourth, how robust are LizTail's benefits across different batch sizes, context lengths, and memory bandwidths? Finally, can the
LizTail Mapper efficiently navigate the enlarged configuration space?

\subsection{Experimental Methodology}
\label{sec:methodology}

\subsubsection{Workloads}
We evaluate both CNN inference and LLM decoding to cover a broad range of GEMM/GEMV shapes. For CNNs, we use ResNet-50 \cite{kansal2024resnet}, MobileNetV3-Large \cite{howard2019searching}, and EfficientNet-B0 \cite{tan2019efficientnet}. ResNet-50 represents conventional CNNs with a relatively high fraction of regular GEMMs, whereas MobileNetV3-Large and EfficientNet-B0 expose more irregular and narrow GEMMs due to their lightweight depthwise/pointwise structures and varying channel dimensions.

For LLMs, we evaluate GPT-2 XL \cite{radford2019language}, LLaMA3 8B \cite{llama3}, and Qwen3-30B-A3B \cite{qwen3_30b_a3b_hf}. These models progressively cover conventional multi-head attention (MHA), grouped-query attention (GQA), and GQA combined with mixture-of-experts (MoE), thereby exposing different projection, FFN, and attention-related GEMM/GEMV shapes. Table~\ref{tab:workloads} summarizes the evaluated model
configurations. Unless otherwise stated, LLM decoding uses batch size $bs=1$ and context length $ctx=1$K. We separately sweep both
parameters in Section~\ref{sec:sensitivity}.

Our main LLM evaluation focuses on autoregressive decoding because, as shown in Section~\ref{WorkloadCharacterizationandDesignMotivation}, small-batch decoding
contains substantially more skinny GEMMs and GEMV-like operations than prefill and therefore directly stresses the array-shape
adaptability targeted by LizTail.

\begin{table}[htbp]
    \centering
    \caption{Evaluated neural-network workloads. $H$ and $F$
    denote the hidden and FFN/intermediate dimensions, respectively;
    $Q$ and $KV$ denote the numbers of query and key/value heads.}
    \label{tab:workloads}
    \resizebox{\columnwidth}{!}{
    \begin{tabular}{llrrrr}
        \toprule
        Type & Model & Layers & Size & $(H,F)$ & $(Q,KV)$ \\
        \midrule
        CNN & ResNet-50          & 50 & 25.6M & -- & -- \\
        CNN & MobileNetV3-Large  & 15 & 5.48M & -- & -- \\
        CNN & EfficientNet-B0    & 16 & 5.29M & -- & -- \\
        \midrule
        LLM & GPT-2 XL           & 48 & 1.5B  & $(1600,6400)$  & $(25,25)$ \\
        LLM & LLaMA3 8B          & 32 & 8B    & $(4096,14336)$ & $(32,8)$ \\
        LLM & Qwen3-30B-A3B      & 48 & 30B   & $(2048,6144)$  & $(32,4)$ \\
        \bottomrule
    \end{tabular}}
\end{table}

\subsubsection{Hardware Configuration and Baselines}

Table~\ref{tab:hw_config} summarizes the default LizTail
configuration. The physical compute array contains $128\times128$
PEs and is partitioned into 32 $128\times4$ segments. The design
operates at 700~MHz and uses 2~MB CEB and 2~MB SIB capacity.
Unless otherwise specified, the off-chip memory bandwidth is
819~GB/s.

\begin{table}[htbp]
    \centering
    \caption{Default LizTail hardware configuration.}
    \label{tab:hw_config}
    \begin{tabular}{lr}
        \toprule
        Parameter & Value \\
        \midrule
        Physical PE array / Array segment& $128\times128$ / $128\times4$\\        
        Number of segments / Frequency& 32 / 700 MHz\\        
        Technology / Memory bandwidth& TSMC 28 nm / 819 GB/s\\
        CEB capacity / SIB capacity& 2 MB / 2 MB\\           
        \bottomrule
    \end{tabular}
\end{table}

We compare LizTail against four representative systolic-array architectures. \textbf{TPU-like} uses a fixed $128\times128$
two-dimensional array and serves as the conventional fixed-array baseline. \textbf{Gemmini} retains a fixed array topology while
supporting more flexible dataflows. \textbf{ReDas} serves as the fine-grained shape-and-dataflow adaptation baseline. Its offline mapper selects a logical shape, dataflow, buffer allocation, tiling, and loop order for each GEMM, after which the array and multi-mode buffers are reconfigured at runtime before execution. \textbf{SARA} exposes a broader scale-up/scale-out configuration space and represents a highly flexible reconfigurable design.

To avoid conflating architectural benefits with compute capacity, all architectures use the same $128\times128$-equivalent PE budget, 700-MHz target frequency, and default off-chip bandwidth.
Importantly, we do not force LizTail's mapping policy onto the baselines. Each architecture is independently mapped within its
native legal configuration space, including the array organizations, dataflows, tiling choices, and loop orders supported by that design.

\subsubsection{Simulation and Hardware Modeling}
Our evaluation combines architecture-level performance simulation, RTL-level hardware implementation, SRAM characterization, and the
offline mapper described in Section~\ref{sec:mapper}. 
SCALE-Sim-v3 \cite{raj2025scale} provides the base systolic execution cycles, while LizTail-specific turn-back, buffer-service, distributed-execution, and reduction overheads are added using the model in Section \ref{sec:cost_model}.

The LizTail PE array, reconfiguration logic, and control datapaths are implemented in Verilog HDL. SRAM structures are generated
using a memory compiler, and logic area and power are obtained using Synopsys Design Compiler under the same TSMC 28-nm
technology setting. The power analysis uses switching activity derived from simulation traces. Unless otherwise stated, the
hardware results use a 0.81-V, $125^\circ$C synthesis corner.

\subsubsection{Metrics}
We report execution latency, speedup, PE utilization, energy-delay product (EDP), area-delay product (ADP), and mapper
search time. Speedup is normalized to the TPU-like baseline, where higher is better. EDP and ADP are defined as
\begin{equation}
    \mathrm{EDP}=E\times T,
    \qquad
    \mathrm{ADP}=A\times T,
\end{equation}
where $E$, $A$, and $T$ denote energy, silicon area, and execution latency, respectively. In Figures~\ref{fig:edp-norm} and
\ref{fig:adp-norm}, EDP and ADP are normalized to LizTail, such that larger values indicate higher cost relative to LizTail.

We report PE utilization over the physical compute resources to capture structural underutilization rather than only the occupancy
of PEs that remain enabled after trimming.

\subsection{Overall Performance and Efficiency}
\label{sec:overall}

\begin{figure}[htbp]

    \centering
    \includegraphics[width=\linewidth]{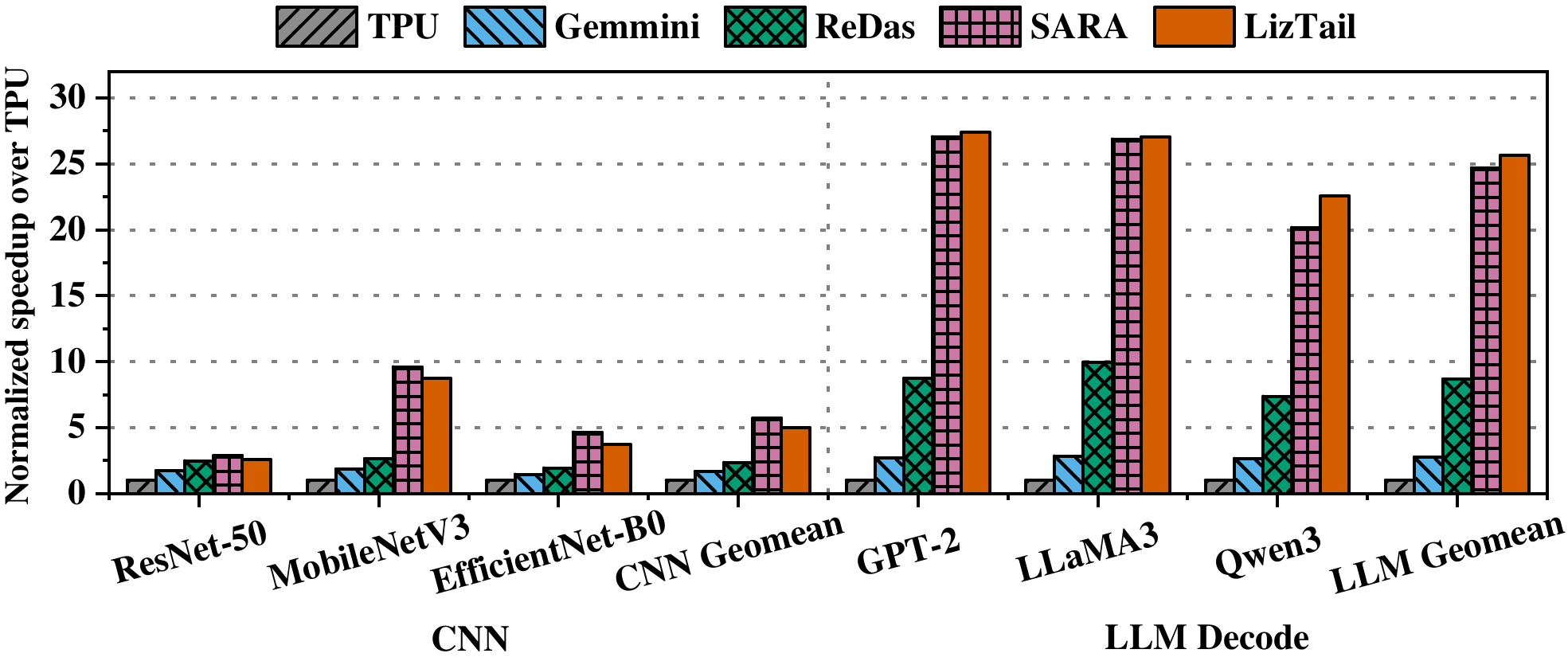}
    \caption{Normalized speedup over the TPU-like baseline across evaluated workloads.}
    \label{fig:norm-speedup}
\end{figure}




Figure~\ref{fig:norm-speedup} compares end-to-end accelerator execution performance across the six workloads. LizTail provides the largest benefits on small-batch LLM decoding, where fixed arrays experience the most severe shape mismatch. Relative to the TPU-like baseline, LizTail achieves $27.39\times$, $27.04\times$, and $22.58\times$ speedups on GPT-2 XL \cite{openai_gpt2xl_hf}, LLaMA3 8B \cite{llama3}, and Qwen3-30B-A3B \cite{qwen3_30b_a3b_hf}, respectively, corresponding to a geometric mean of $25.67\times$.

ReDas \cite{han2024redas} is the strongest baseline for isolating the benefit of spatial parallelism because it already constructs elongated logical arrays beyond the physical side length and selects among OS, WS, and IS. Nevertheless, LizTail is \(3.14\times\), \(2.72\times\), and \(3.07\times\) faster on GPT-2 XL, LLaMA3 8B, and Qwen3-30B-A3B, respectively. ReDas executes each GEMM through one reshaped logical array, whereas LizTail can decompose partitionable operations across concurrent subarrays and locally extend each partition when its assigned workload remains shape-mismatched.
This gap indicates that executing each GEMM through one reshaped logical array is insufficient for many decoding workloads: some layers benefit more from decomposing the computation into parallel subtasks, while others benefit from combining such parallelism with local extension.

CNN gains are smaller but remain substantial. LizTail provides $2.60\times$, $8.74\times$, and $3.70\times$ speedups over the
TPU-like baseline on ResNet-50, MobileNetV3-Large, and EfficientNet-B0, respectively, for a geometric mean of $5.02\times$. MobileNetV3-Large benefits the most because its lightweight operators expose a larger number of narrow and small GEMMs for which a monolithic $128\times128$ organization wastes a substantial fraction of the available compute resources.

\begin{figure}[htbp]
    \centering
    \includegraphics[width=\linewidth]{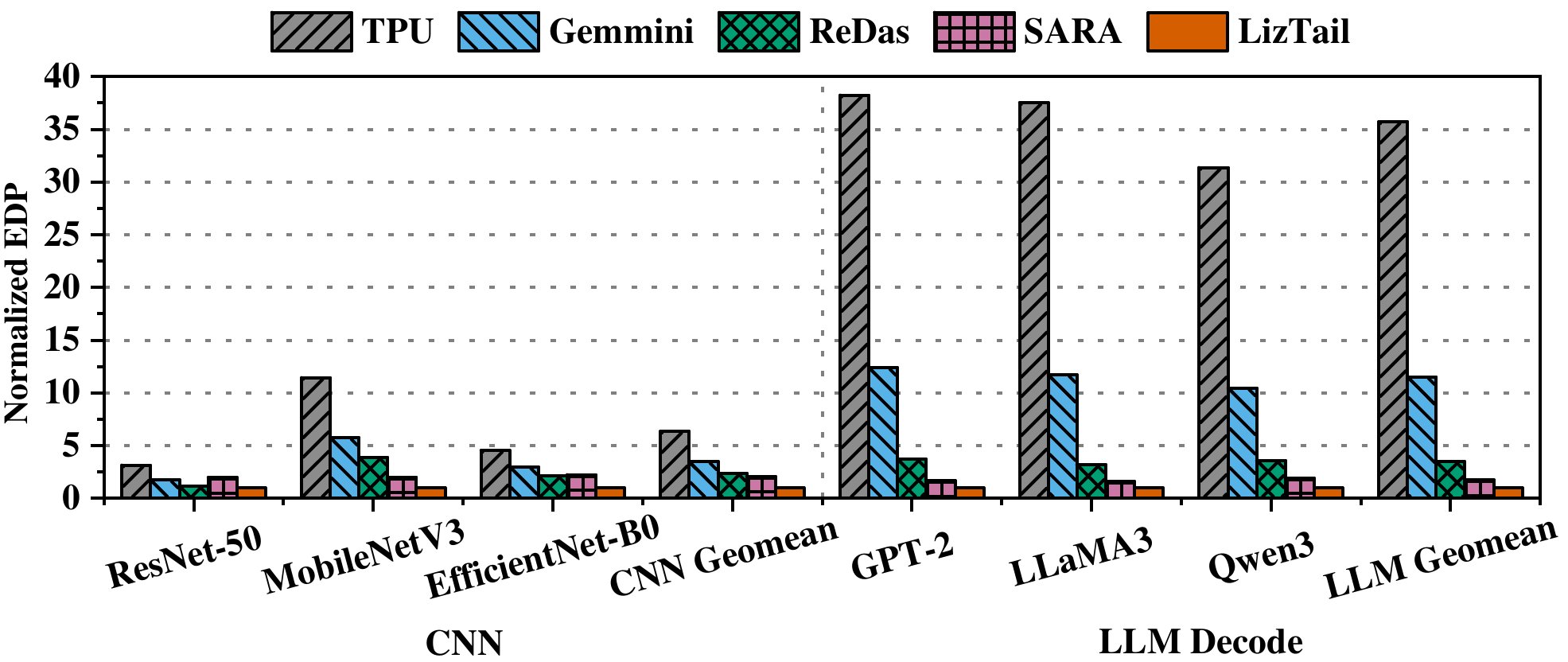}
    \caption{Normalized EDP across evaluated workloads.}
    \label{fig:edp-norm}
\end{figure}

Performance alone, however, does not capture the cost of reconfiguration. Figure~\ref{fig:edp-norm} reports normalized EDP.
LizTail achieves the lowest EDP across all six evaluated workloads.
Relative to the TPU-like baseline, it reduces EDP by $3.13\times$, $11.43\times$, and $4.52\times$ on the three CNNs,
and by $38.23\times$, $37.50\times$, and $31.37\times$ on the three LLMs. Relative to ReDas, the corresponding EDP reductions
are $1.15\times$--$3.87\times$ for CNNs and $3.17\times$--$3.68\times$ for LLM decoding.

SARA \cite{samajdar2022self} can achieve competitive latency through fine-grained scale-out configurations. This flexibility, however, has different area and energy implications: dense SRAM banking increases silicon footprint, while the bypass and operand-distribution network increases array and communication power. Consequently, its latency advantage does not translate directly into comparable EDP or ADP improvements.

\begin{figure}[htbp]
    \centering
    \includegraphics[width=\linewidth]{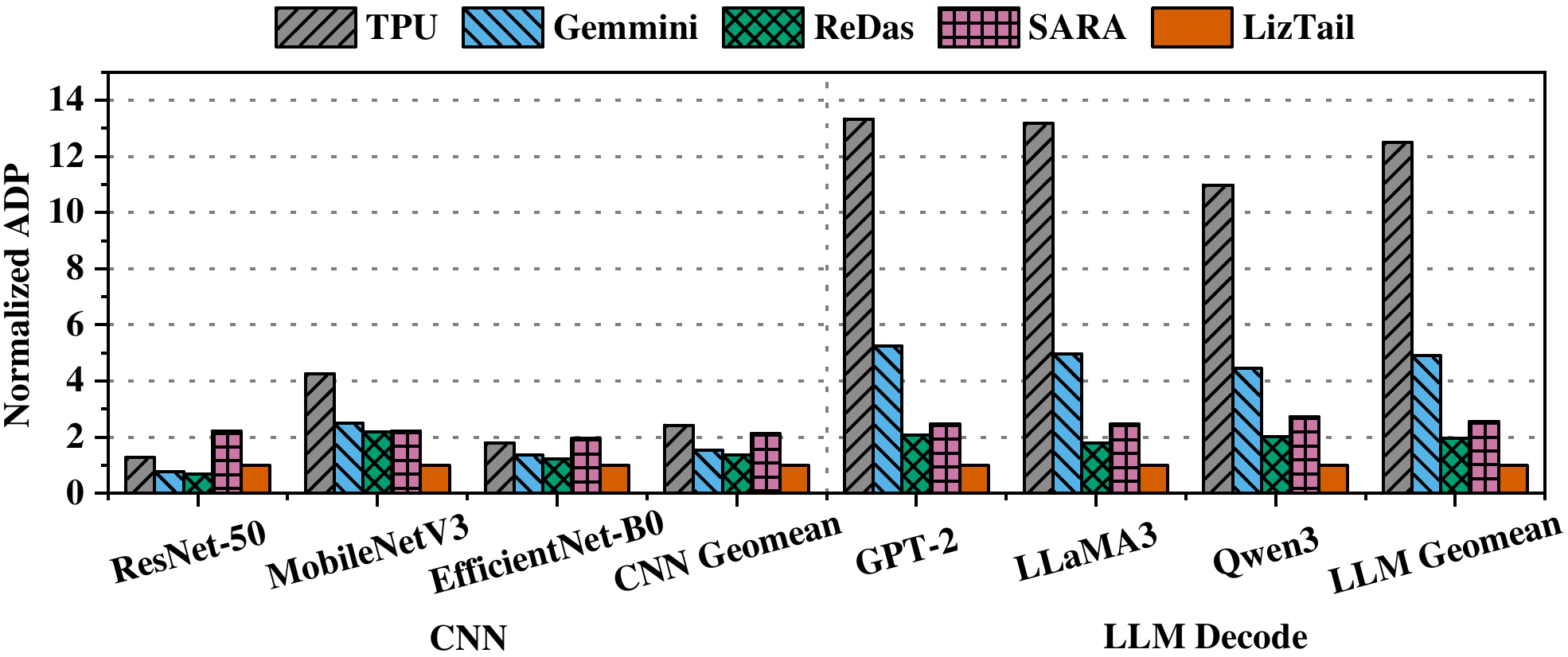}
    \caption{Normalized ADP across evaluated workloads.}
    \label{fig:adp-norm}
\end{figure}

Figure~\ref{fig:adp-norm} further captures the interaction between performance and silicon cost. Relative to the TPU-like baseline,
LizTail reduces ADP by $10.67\times$--$12.94\times$ across the three LLM decoding workloads. Compared with SARA, LizTail
reduces ADP by $2.40\times$, $2.39\times$, and $2.66\times$ on GPT-2 XL, LLaMA3 8B, and Qwen3-30B-A3B, respectively.

These results highlight the central design point of LizTail: performance does not improve simply by maximizing
reconfigurability. Instead, exposing shape adaptation and spatial parallelism through a constrained segment-level substrate provides
a better overall performance --energy--area trade-off.

\subsection{Understanding LizTail's Gains}
\label{sec:understanding}

We next examine why LizTail improves performance. We first measure PE utilization, then isolate the contribution of shape extension and distributed parallelism, and finally examine which modes are actually selected by the mapper.

\subsubsection{PE Utilization}
\begin{figure}[htbp]
    \centering
    \includegraphics[width=\linewidth]{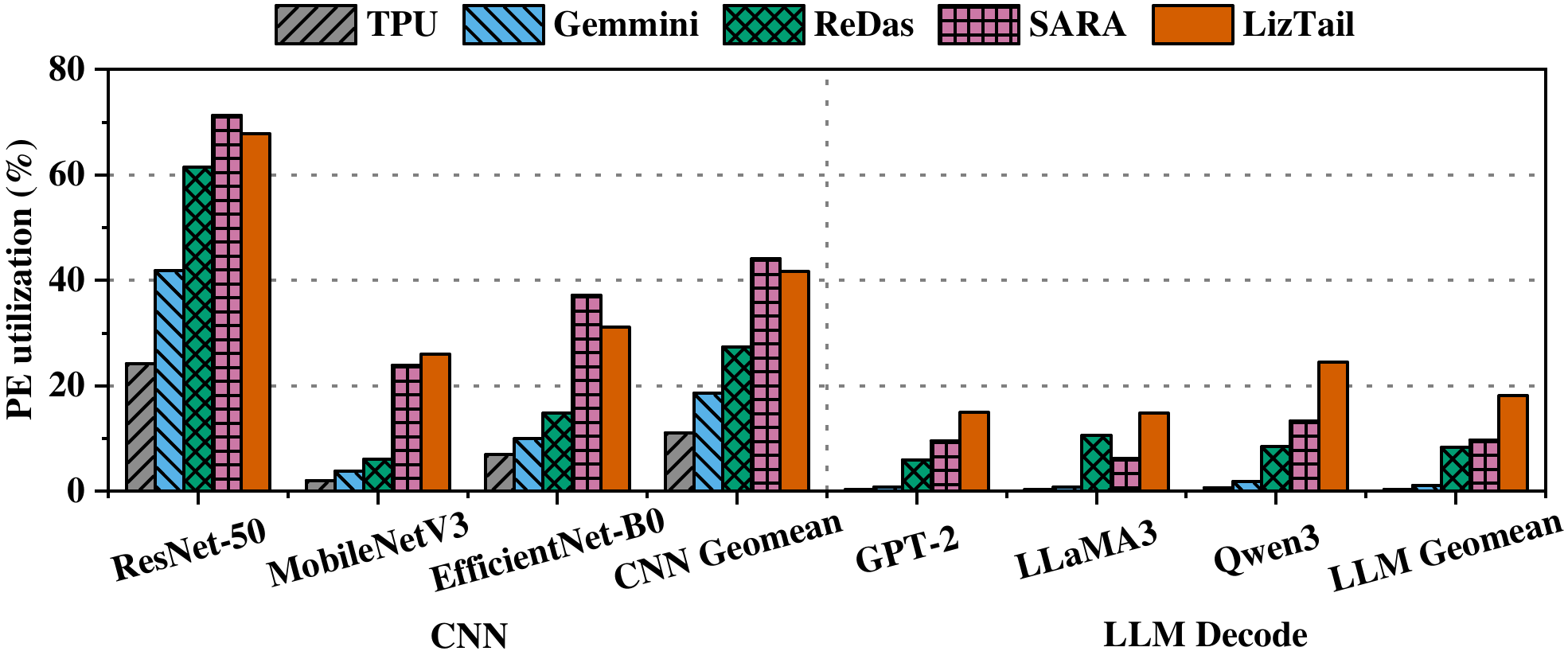}
    \caption{PE utilization across evaluated workloads.}
    \label{fig:pe-utilization}
\end{figure}

Figure~\ref{fig:pe-utilization} shows that LizTail's performance gains closely track improved utilization of the physical compute array.
On ResNet-50, LizTail reaches $67.87\%$ utilization, compared with $24.22\%$ for TPU-like, $41.84\%$ for Gemmini, and $61.44\%$ for
ReDas. SARA reaches $71.29\%$, illustrating that aggressive reconfigurability can further increase instantaneous utilization,
but Section~\ref{sec:hardware_cost} shows that this flexibility comes with substantially higher area cost.

The utilization gap is more pronounced on workloads with narrow GEMMs. LizTail achieves $26.01\%$ and $31.10\%$ utilization on
MobileNetV3-Large and EfficientNet-B0, respectively, despite the large fraction of layers whose dimensions poorly match the native
$128\times128$ geometry.

Small-batch LLM decoding remains intrinsically difficult to parallelize, so absolute utilization is lower than in CNN inference.
Nevertheless, LizTail reaches $14.91\%$, $14.85\%$, and $24.51\%$ on GPT-2 XL, LLaMA3 8B, and Qwen3-30B-A3B, respectively, while
the fixed-array baselines remain in the low-single-digit range.
LizTail therefore converts a larger fraction of the same physical MAC budget into useful work even when the workload is highly
GEMV-like.

\begin{figure}[htbp]
    \centering
    \includegraphics[width=\linewidth]{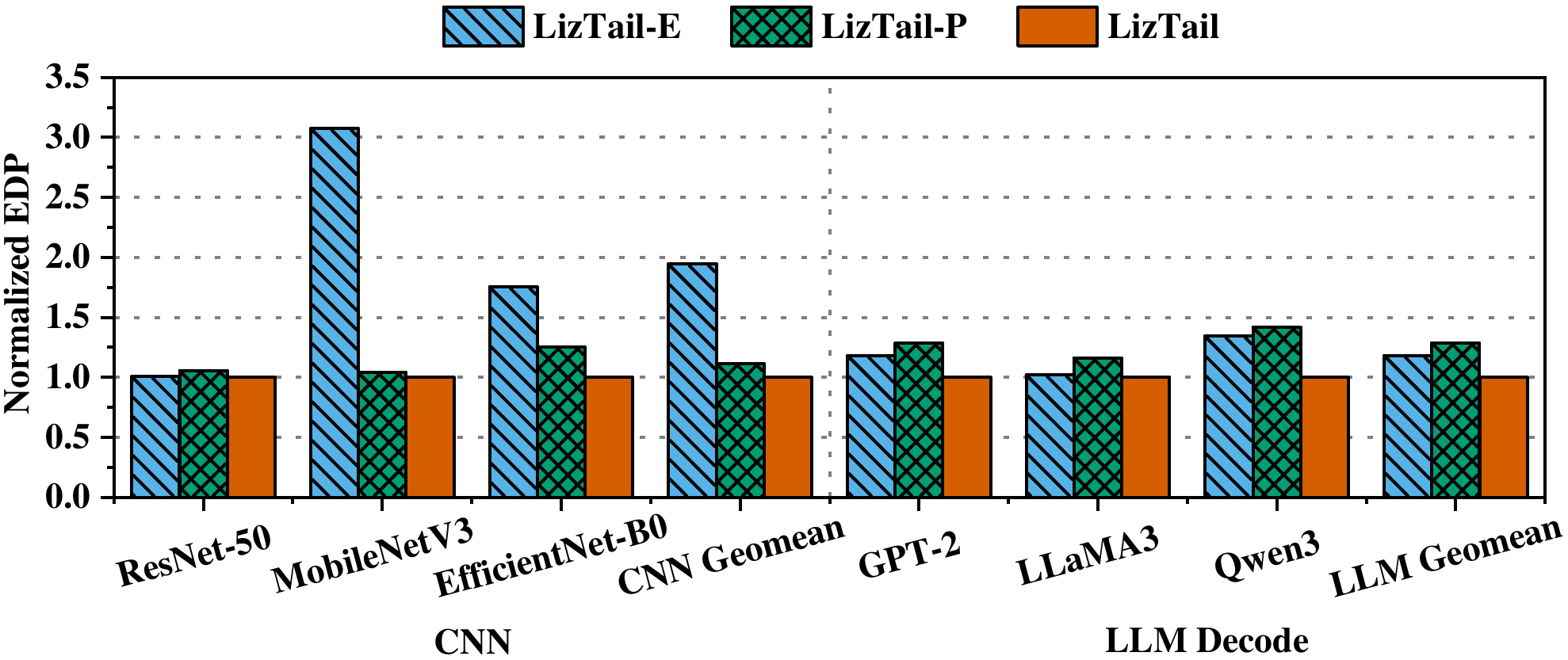}
    \caption{Ablation study of LizTail execution modes.}
    \label{fig:ablation-edp}
\end{figure}

\subsubsection{Mode Ablation}
To isolate the contribution of the two primary adaptation mechanisms, we construct two restricted LizTail variants.
\textbf{LizTail-E} retains Trimming and Extension Modes but disables distributed parallelism, whereas \textbf{LizTail-P}
retains Trimming and Distributed-Parallel Modes but disables cross-segment extension. The full LizTail design enables all three
modes. All variants use the same underlying PE budget and mapper objective.

Figure~\ref{fig:ablation-edp} reports average EDP normalized to the full LizTail. For CNNs, LizTail-E, LizTail-P, and full LizTail
achieve normalized EDPs of $1.9, 1.2$, and $1.0$, respectively.
Distributed parallelism therefore contributes more strongly to CNN efficiency: many lightweight CNN operators can be decomposed into
independent subproblems, while extension alone cannot fully exploit this task-level parallelism.


\begin{figure}[!htbp]
\centering
\begin{subfigure}[t]{0.23\textwidth}
\centering
\includegraphics[width=\linewidth]{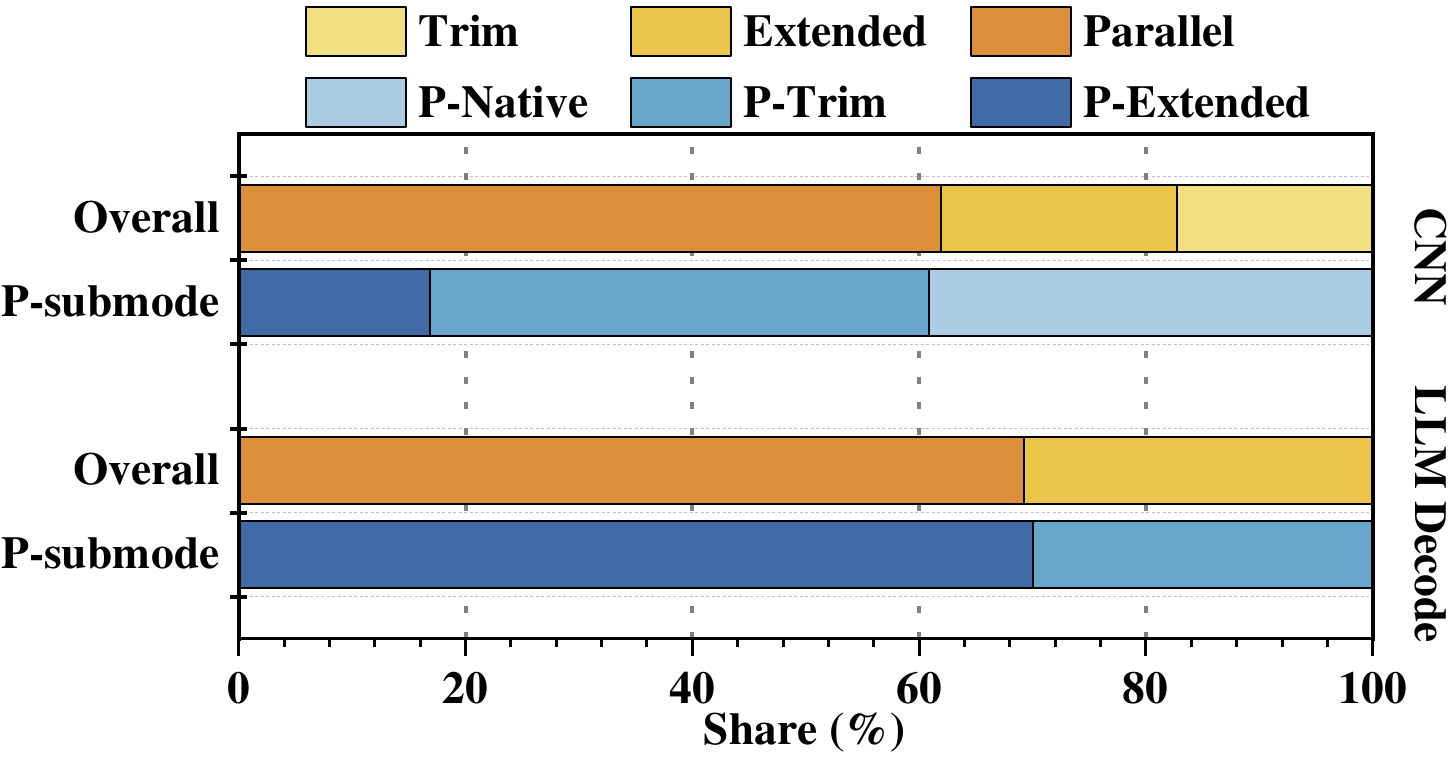}
\subcaption{}
\label{fig:mode_breakdown}
\end{subfigure}
\begin{subfigure}[t]{0.24\textwidth}
\centering
\includegraphics[width=\linewidth]{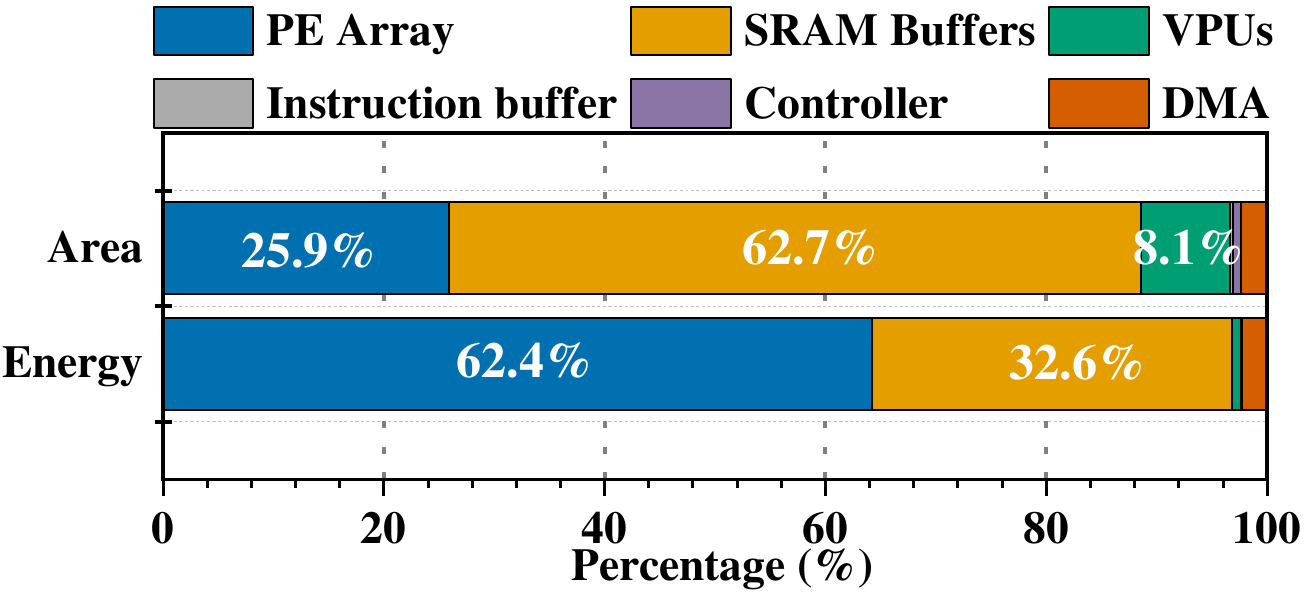}
\subcaption{}
 \label{fig:area_energy_breakdown}
\end{subfigure}
\caption{LizTail mapping decisions and hardware cost. (a) Distribution of execution modes and P-Mode configurations selected by the LizTail Mapper across CNN and LLM decoding workloads. (b) Breakdown of LizTail’s area and energy consumption by hardware component.}
\label{fig:batch_ctx}
\end{figure}

The behavior changes for LLM decoding. LizTail-E, LizTail-P, and full LizTail obtain normalized EDPs of $1.3, 1.4$, and $1.0$,
respectively. Elongated projection and FFN operations benefit substantially from long logical paths, but extension alone misses
parallelizable output partitions. Conversely, parallel execution without local extension leaves each subarray vulnerable to the same
shape mismatch. The full design performs best because it can select between extension, parallel execution, and their composition on a per-workload basis.

\subsubsection{Which Modes Does the Mapper Select?}
Figure~\ref{fig:batch_ctx}(a) reports the execution-mode distribution selected by the LizTail Mapper. For CNN workloads,
Trimming, Extension, and Distributed-Parallel Modes account for $17.3\%$, $20.8\%$, and $61.9\%$ of the mapped operations,
respectively. Within P-Mode, P-Native, P-Trim, and P-Extended account for $39.1\%$, $44.1\%$, and $16.9\%$, respectively.

LLM decoding exhibits a different distribution: standalone Trimming is not selected, while standalone Extension and P-Mode
account for $30.7\%$ and $69.3\%$, respectively. More importantly, P-Mode does not correspond to fixed-shape subarray execution.
Within P-Mode, $30.0\%$ of configurations use P-Trim and $70.0\%$ use P-Extended, while P-Native is not selected.

This result directly supports the workload characterization in Section~\ref{WorkloadCharacterizationandDesignMotivation}. Shape extension and spatial parallelism are not competing alternatives. In many LLM layers, extension is most useful \emph{inside} concurrently executing subarrays, explaining why the composable P-Extended mode is frequently selected and why neither LizTail-E nor LizTail-P matches the complete architecture.

\subsection{Hardware Cost of Reconfiguration}
\label{sec:hardware_cost}

We next quantify the hardware cost required to support LizTail's additional execution modes. All compared designs are implemented
under the same 28-nm synthesis methodology. SRAM structures are generated using the same memory-compiler flow, while the PE array,
controller, and reconfiguration logic are synthesized from Verilog RTL. Dynamic power is estimated using simulation-derived switching activity.


Figure~\ref{fig:batch_ctx}(b) breaks down LizTail's area and energy. SRAM buffers dominate silicon area, accounting for
$62.7\%$, while the PE array contributes $25.9\%$ and the vector processing units contribute $8.1\%$. The energy distribution is
different: the PE array contributes $62.4\%$ of execution energy, whereas SRAM buffers account for $32.6\%.$ This contrast motivates
the constrained SIB design in Section~\ref{sec:sram_buffer}: providing unrestricted operand injection through heavily banked
global buffers would directly amplify the component that already dominates area.

The differentiated PE organization also limits compute-side reconfiguration overhead. The synthesized PE-array logic of LizTail is approximately $10.9\%$ smaller than that of ReDas PE array in our implementation, because bidirectional and endpoint-turning logic is instantiated only at locations that participate in legal LizTail paths rather than replicated uniformly across all PEs.

\begin{figure}[htbp]
    \centering
    \includegraphics[width=\linewidth]{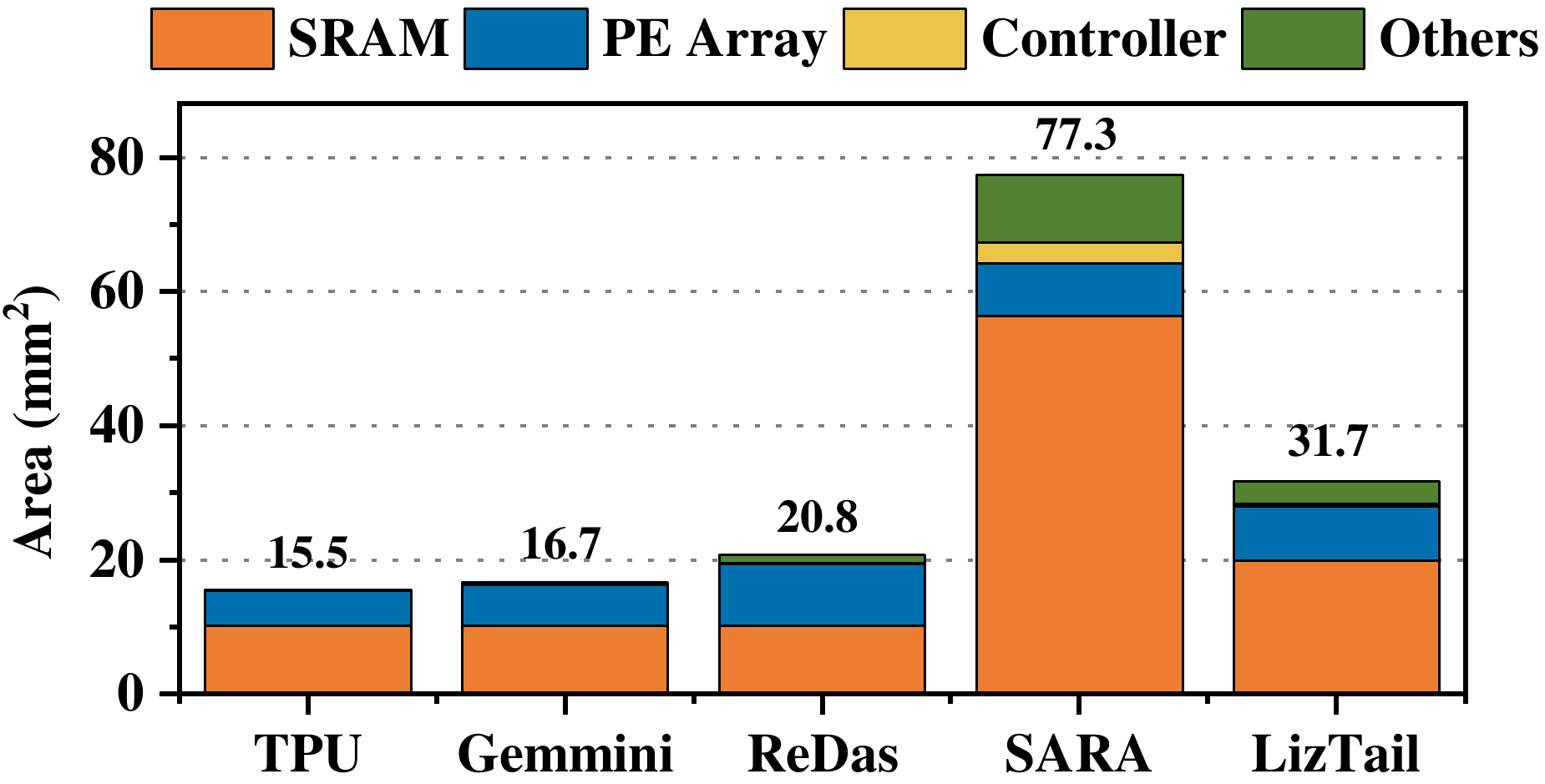}
    \caption{Comparison of area composition across different systolic array architectures.}
    \label{fig:area_comparison_stacked}    
    \vspace{-0.2in}
\end{figure}

Figure~\ref{fig:area_comparison_stacked} compares total area across architectures. TPU-like, Gemmini, ReDas, SARA, and LizTail occupy $15.5, 16.7, 20.8, 77.3$, and $31.7$~mm$^2$, respectively. LizTail therefore costs more silicon than fixed and reshape-only designs, as expected from its additional SIB and distributed-execution support.
The resulting LLM latency reduction nevertheless outweighs this area increase, allowing LizTail to retain lower ADP.

Combined with the ADP results in Section~\ref{sec:overall}, these measurements show that LizTail's performance benefits are not
obtained by simply provisioning the largest reconfiguration fabric. Constraining flexibility to regular segments and predefined ingress points preserves most of the useful execution organizations while avoiding the hardware cost of unrestricted PE-level routing and data delivery.

\subsection{Sensitivity to Inference and Memory Conditions}
\label{sec:sensitivity}
\begin{figure}[!htbp]
\centering

\begin{subfigure}{0.23\textwidth}
\centering
\includegraphics[width=\linewidth]{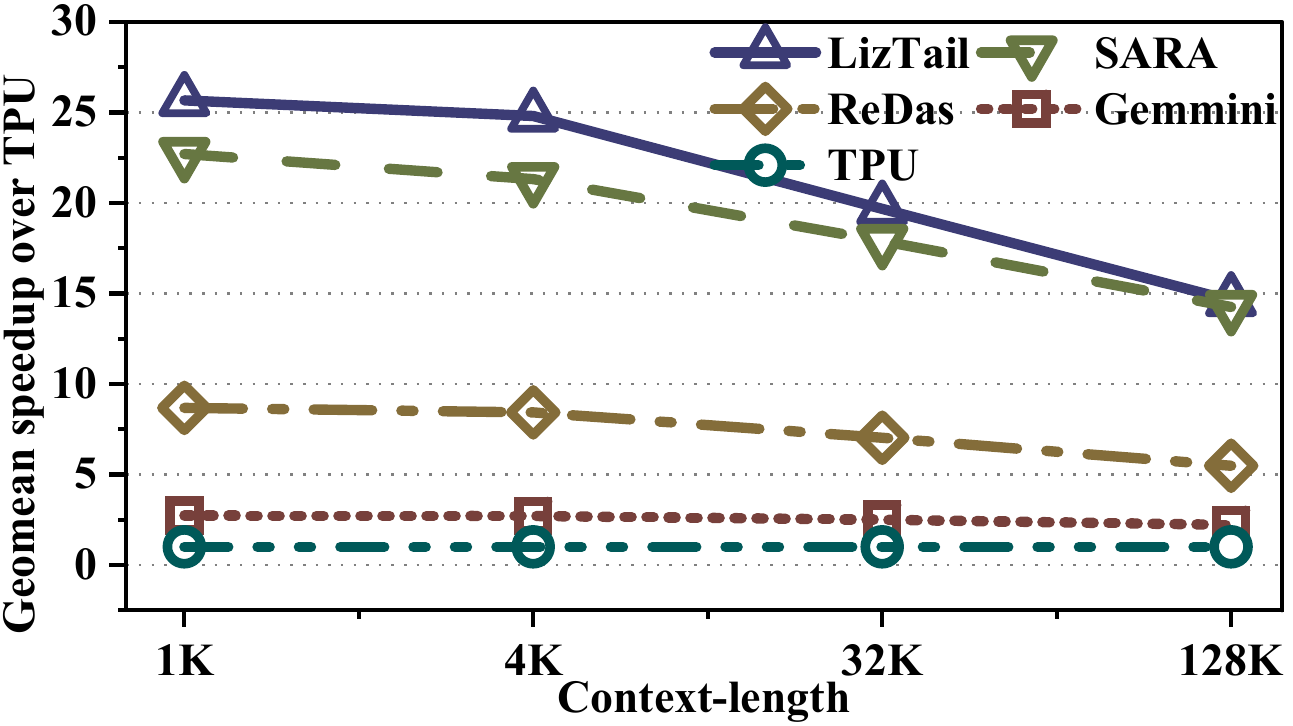}
\caption{Context‑length sweep}
\end{subfigure}
\begin{subfigure}{0.23\textwidth}
\centering
\includegraphics[width=\linewidth]{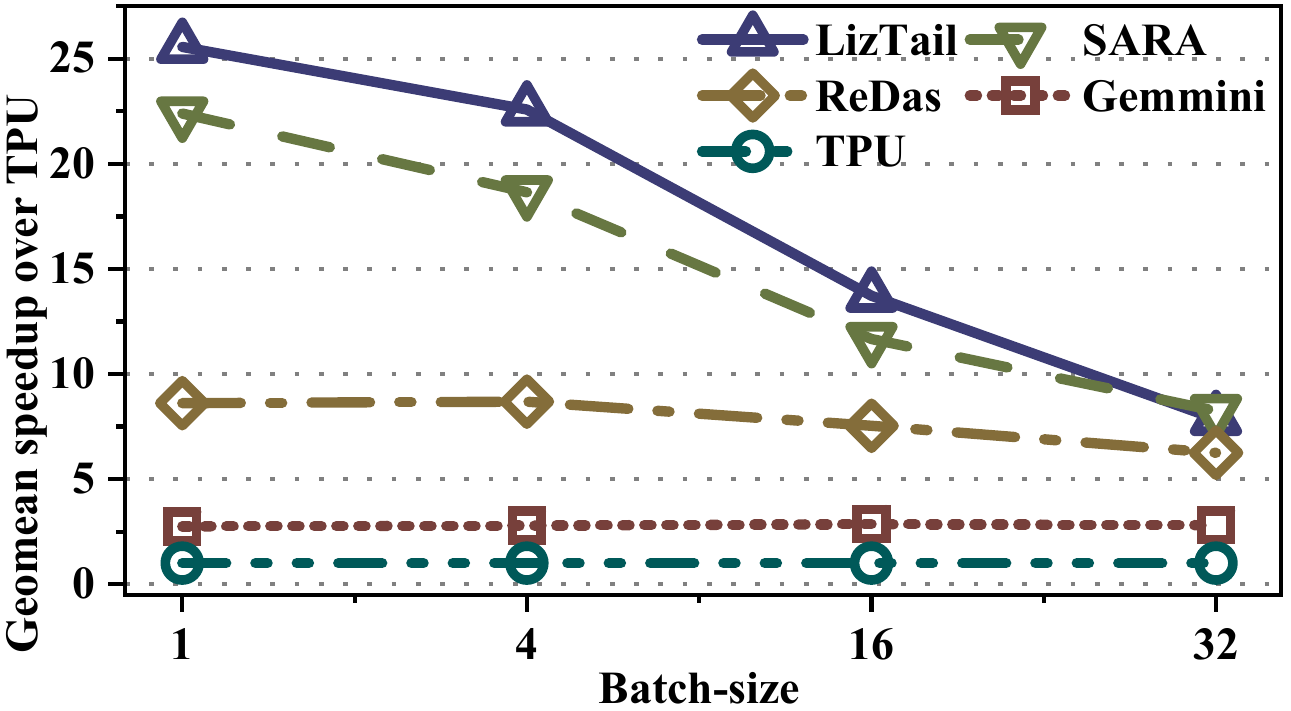}
\caption{Batch‑size sweep}
\end{subfigure}
\caption{Normalized speedup of LLM decoding across different batch sizes and context lengths.}
\label{fig:scan_bs_ctx}
\end{figure}

\begin{figure}[!htbp]
\centering
\begin{subfigure}{0.23\textwidth}
\centering
\includegraphics[width=\linewidth]{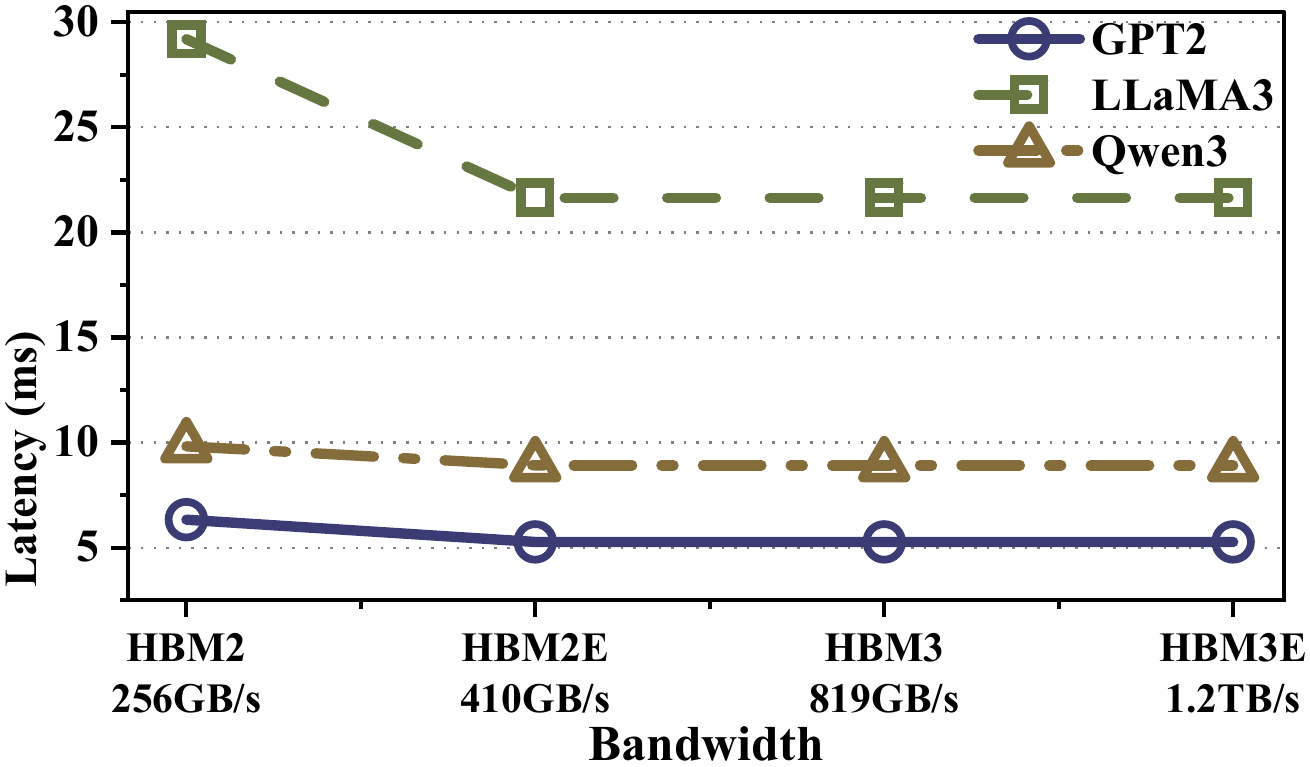}
\caption{Decode (bs=32, ctx=1K)}
\end{subfigure}
\begin{subfigure}{0.23\textwidth}
\centering
\includegraphics[width=\linewidth]{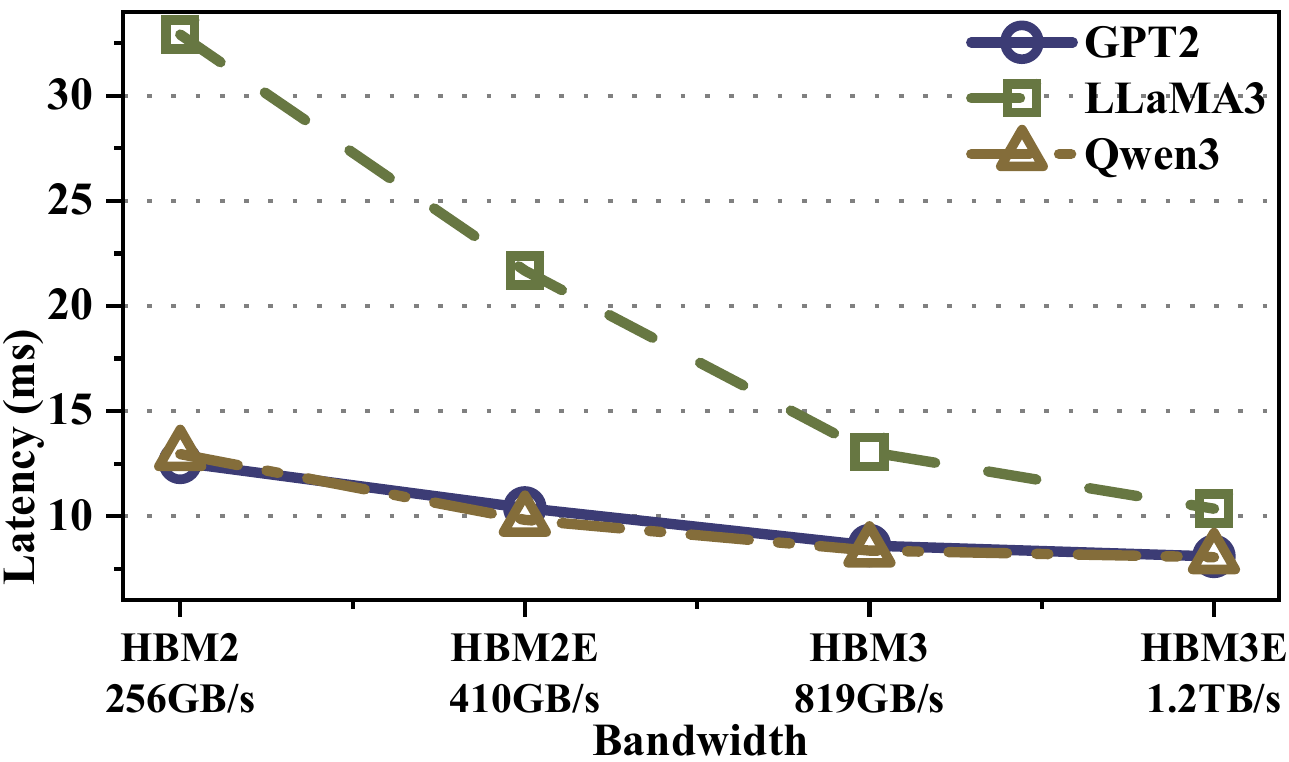}
\caption{Decode (bs=1, ctx=128K)}
\end{subfigure}
\caption{LLM decoding latency under different HBM bandwidth configurations.}
\label{fig:bandwidth}
\end{figure}

Figures~\ref{fig:scan_bs_ctx} and~\ref{fig:bandwidth} evaluate workload and memory sensitivity. Increasing batch size from 1 to 32 reduces LizTail's geometric-mean speedup from $25.57\times$ to $7.86\times$, because the short GEMM dimension grows and fixed arrays become better utilized. Increasing context length from 1K to 128K reduces speedup from $25.57\times$ to $14.09\times$ as KV-cache traffic occupies a larger fraction of per-token latency.

At $bs=32$ and $ctx=1$K, performance largely saturates at 410--819~GB/s, whereas $bs=1$ and $ctx=128$K remains strongly memory-sensitive. LizTail therefore addresses compute-shape inefficiency but does not eliminate the memory bottleneck of long-context decoding.

\subsection{Mapper Quality and Search Overhead}
\begin{figure}[htbp]
    \centering
    \includegraphics[width=\linewidth]{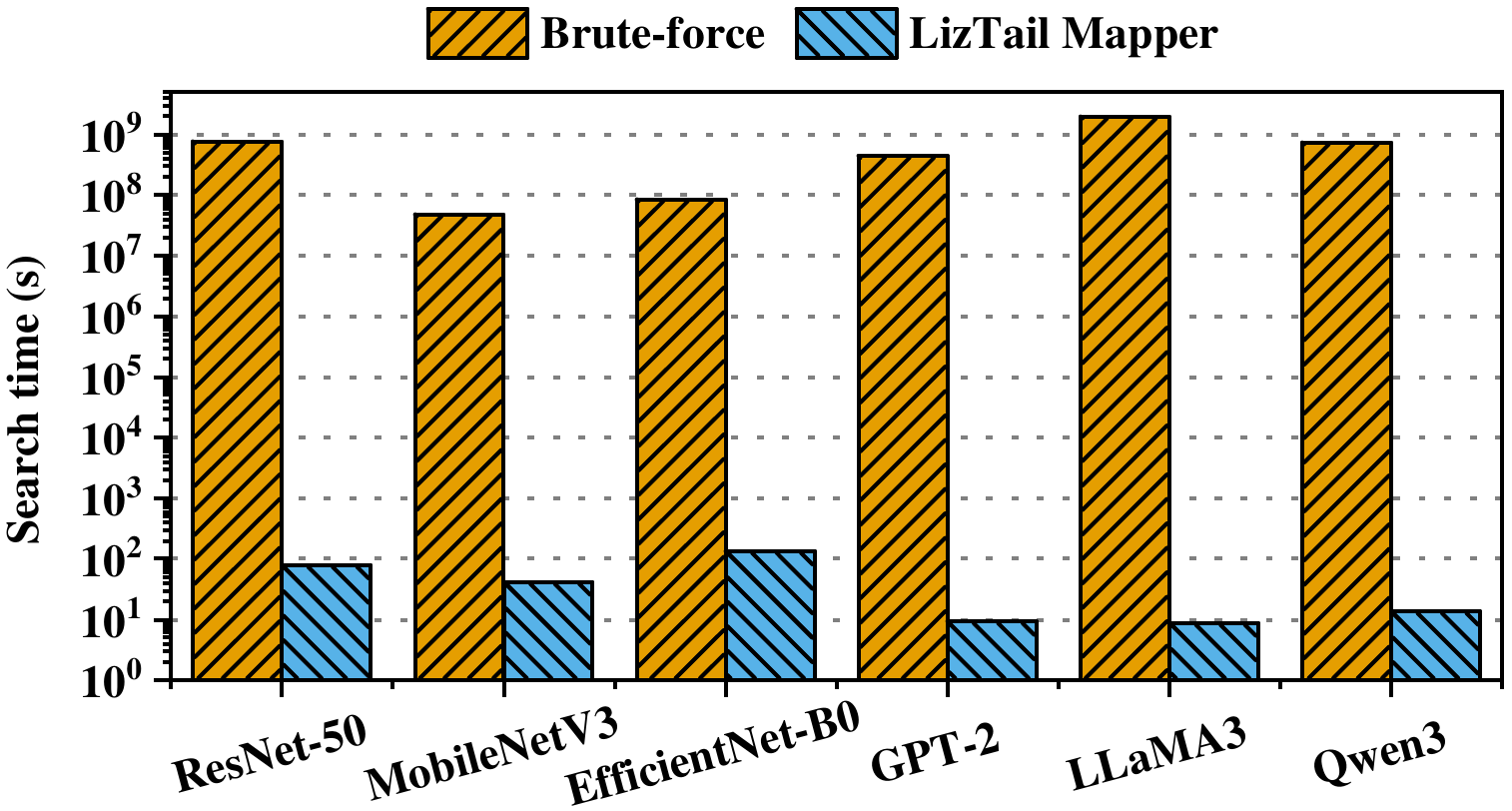}
    \caption{Comparison of search times between LizTail Mapper and brute-force enumeration.}
    \label{fig:mapper_time}
    \vspace{-0.1in}
\end{figure}


LizTail's multi-mode execution space increases compile-time mapping choices. The Python 3.8 mapper operates offline and generates the logical-array, dataflow, tiling, partitioning, and buffer configurations used by SCALE-Sim-v3 \cite{raj2025scale}; its search time is therefore excluded from inference latency. 

Figure~\ref{fig:mapper_time} compares whole-model search time against brute-force enumeration of the complete legal mapping space. Brute-force search requires an estimated ($10^{7}$)--($10^{9}$) seconds, whereas LizTail Mapper completes all evaluated workloads within tens to hundreds of seconds. This corresponds to reductions of up to seven orders of magnitude for CNNs and more than eight orders for LLM decoding. 
On a representative set of tractable GEMM/GEMV instances for which exhaustive enumeration is feasible, LizTail's selected configurations are within $1.11\%$ of the exhaustive optimum on average.

The reduction comes from four complementary mechanisms described in Section~\ref{sec:mapper}: segment-constrained candidate generation, critical-tiling selection, parallel T/E/P-Mode search, and repeated-workload caching. Search time consequently depends on the number of distinct GEMM/GEMV shapes rather than model size. Transformer blocks repeatedly use identical projection and FFN shapes and therefore achieve high cache reuse, whereas the more heterogeneous convolution shapes across CNN stages require additional candidate evaluations.

\vspace{-0.1in}
\section{Related Work}

\textbf{Reconfigurable systolic arrays.}
Gemmini decouples dataflow selection from a largely fixed array topology \cite{genc2021gemmini}. DyNNamic co-designs compact-CNN decomposition with output-stationary reshaping, whereas ReDas chains four physical regions through PE-local roundabout paths to support fine-grained logical shapes under OS, WS, and IS \cite{hanson2022dynnamic,han2024redas}. ReSA instead targets the concurrent execution of multiple tiny DNN tensors \cite{lee2024resa}. At a coarser granularity, Planaria dynamically fissions an array for spatial multi-tenancy, while SARA jointly explores scale-up/scale-out configurations and mappings on a reconfigurable-array substrate \cite{ghodrati2020planaria,samajdar2022self}. These designs establish the benefits of dataflow, shape, and partitioning flexibility, but instantiate routing and operand delivery at different granularities. LizTail restricts reconfiguration to regular segments and predefined endpoints, thereby supporting trimming and long-path extension without unrestricted PE-level routing.

\textbf{Scale-out and hybrid organizations.}
SOSA co-designs pod granularity, interconnect, and tiling for fixed multi-pod accelerators, and SAVector similarly uses multiple small arrays to expose tile-level parallelism \cite{yuzuguler2023scale,choi2024savector}. SISA is the closest contemporary design: it partitions a square array into horizontal slabs that execute independently or fuse, up to the original array extent, under an output-stationary dataflow \cite{altamura2026sisa}. LizTail additionally constructs logical dimensions beyond the physical array through 180-degree turn-back paths and composes such extension within concurrently executing subarrays. Its T/E/P-Modes share one segment-reconfigurable substrate and support OS, WS, and IS through differentiated PEs and segment-aware data delivery.

\textbf{LLM inference acceleration.}
DeepSpeed Inference and vLLM optimize model execution, batching, and KV-cache management \cite{aminabadi2022deepspeed,kwon2023efficient}; FlashDecoding++ and FlashInfer optimize flat-GEMM and attention kernels \cite{hong2024flashdecoding++,ye2025flashinfer}. AIF and DOTS reduce data movement through in-flash processing and DRAM-PIM, respectively, while DESA specializes systolic dataflow for Transformers \cite{lee2025aif,park2025dots,wang2025desa}. These techniques are complementary to LizTail: they optimize serving, kernels, or memory placement, whereas LizTail adapts the underlying systolic organization to both matrix geometry and operator parallelism across CNN and LLM inference.

\vspace{-0.1in}
\section{Conclusion}
We presented LizTail, a segment-reconfigurable systolic array that unifies shape adaptation and spatial parallelism through a common constrained substrate. Endpoint-based extension, composable subarray execution, segment-aware data delivery, and a hardware-constrained mapper enable LizTail to adapt to diverse CNN and LLM GEMM/GEMV workloads without unrestricted PE-level routing. Our results show that combining these complementary execution modes provides substantial performance and energy-efficiency gains with bounded hardware cost.

\bibliographystyle{ACM-Reference-Format}
\bibliography{sample-base}










\end{document}